# Emerging and established topics in drone research: Citation impact and knowledge flows across China, the United States, the EU, Ukraine, and Russia (2020–2025)

**Corresponding author:**
Myrolava Hladchenko
Centre for R&D Monitoring (ECOOM), University of Antwerp, Belgium
hladchenkom@gmail.com

## Abstract

This study examined emerging and established topics in drone research, focusing on citation impact and knowledge flows across China, the United States, the EU, Ukraine, and Russia between 2020 and 2025 using OpenAlex bibliographic data. The findings revealed that drone-related science is characterised by growing geopolitical asymmetries in scientific production, citation concentration, and international knowledge exchange. In particular, China increasingly dominated scientific production, fractional authorship contribution, and domestic citation circulation. In contrast, the United States and EU countries maintained comparatively more internationally distributed citation structures. However, China-affiliated publications became increasingly integrated into global citation networks, particularly through growing citation exchange with the United States and European countries.

Notably, the interpretation of authorship and citation patterns was complicated by the high proportion of publications with unidentified affiliations, which reached 50% in 2025 within weak-signal topics. These findings underscore the importance of developing comprehensive national Research Organisation Registries (RORs).

Although China demonstrated a citation advantage, this was partly driven by high internal domestic citation concentration rather than exclusively by global integration. Moreover, China still imported proportionally more knowledge from the EU-14 and the United States than it exported, with this asymmetry increasing over time. EU-14 countries maintained the strongest citation impact in weak-signal topics, suggesting a more prominent role in shaping emerging research directions. At the same time, China-affiliated publications cited the United States more frequently than the EU-14 in both strong- and weak-signal topics, with this pattern being particularly pronounced in weak-signal areas.

## Introduction

Tracking the evolution of scientific fields (Wieczorek et al., 2021; Lu et al., 2022) has become increasingly important in an era of rapid technological change, in which both established research areas and nascent directions can have strategic and socioeconomic implications (Park & Kim, 2024). Anticipating changes in the technological landscape is crucial for governments and businesses seeking to plan their research and development (R&D) efforts strategically in order to maintain innovation leadership and competitive advantage (Viet & Gneushev, 2021; Ye et al., 2014; Small et al., 2014). Topic modelling has emerged as a key approach for tracking change in scientific fields (Jiang et al., 2025).

Because scientific fields are dynamic rather than stable, research topics vary substantially in their maturity, scale, growth dynamics, and degree of novelty. Treating them as equivalent risks obscuring key distinctions between established areas and emerging, exploratory directions (Small et al., 2014; Rotolo et al., 2015; Lu et al., 2022). To address this limitation, recent scientometric studies operationalise emergence primarily through growth dynamics, leading to approaches that distinguish between strong-signal (established, rapidly expanding) and weak-signal (nascent, exploratory) topics (Kölbl & Grottke, 2020; Park & Kim, 2021; Zhao et al., 2024).

Against this backdrop, the global scientific system has undergone a profound transformation over the past two decades. China has emerged as a scientific superpower, in part due to international knowledge acquisition, large-scale public investment, and strategic prioritisation of key technologies (Hannas et al., 2020; *The Economist*, 2024). Western governments have increasingly expressed concerns about China's growing technological capabilities and the role of international research collaboration and technology transfer in strengthening its military and dual-use capacities (von der Leyen, 2023). In response, the European Union has elevated research security to a strategic priority within its European Economic Security Strategy, introducing measures to assess risks related to critical technologies and to limit cooperation in sensitive areas (European Commission & High Representative of the Union for Foreign Affairs and Security Policy, 2023; Council of the European Union, 2024). In late 2025, the European Commission announced the exclusion of Chinese universities, including the so-called "Seven Sons of National Defence," from participating in roughly half of Horizon Europe calls (Matthews, 2025). In turn, EU member states have implemented national research security policies (Rüland et al., 2025; Lundin & Shih, 2025; Houttekier et al., 2025; Brouns, 2023).

These policy shifts are likely to reshape international collaboration networks and thematic specialisation in strategically sensitive research areas, making them directly relevant for scientometric analyses of topic evolution and impact.

Striking the right balance between research security and openness to international collaboration has become a key concern for universities, research funders, and governments (James et al., 2025). European Union (EU) member states increasingly view international research collaboration as a potential gateway for foreign interference and knowledge leakage. At the same time, international collaboration benefits both individual scholars and national research systems by providing access to resources and expertise necessary for the creation of new knowledge and technologies (Wang et al., 2024). International research collaboration has increased globally over time (Abramo et al., 2019; Gazni et al., 2012) in response to the growing complexity of science, as tackling major problems requires abundant funding, expensive research infrastructure, and multidisciplinary research teams (Adams, 2013). In scholarly publishing, the citation advantage of international collaboration is well documented. Numerous studies show that internationally co-authored publications tend to achieve higher citation impact than domestically produced papers, reflecting broader knowledge integration, increased visibility, and access to complementary expertise across national boundaries (Glänzel & Schubert, 2001; Glänzel, 2001; Persson et al., 2004; Larivière et al., 2015; De Moya-Anegón et al., 2018).

Within the current geopolitical context, drone technologies (unmanned aerial vehicles, UAVs) represent a particularly salient research domain. Drone research spans both civilian and military applications and intersects closely with artificial intelligence, autonomous navigation, sensing, and data processing—areas that feature prominently in recent assessments of emerging and critical technologies (European Commission, 2023). The dual-use nature of drones makes them a focal point of security-oriented research policy debates and an ideal case for examining how geopolitical tensions shape scientific collaboration, topic evolution, and citation impact.

Accordingly, this study addresses two research questions:

RQ1. How did the drone research evolve between 2020 and 2025?

RQ2. How did China, the United States, the EU, Ukraine, and Russia differ in knowledge flows and citation impact in strong- and weak-signal research topics in drone research?

To address these questions, the study uses the fractional authorship contributions across countries, citation impact and Knowledge Exchange Balance (KEB) as key indicators of scientific production and influence. The study is based on large-scale bibliometric data from the OpenAlex Multiobs project and employs transformer-based topic modelling combined with signal classification techniques (Ebadi et al., 2025) to identify and characterise evolving research themes and their temporal dynamics.

## Strong- and weak-signal topics

Topic modelling allows conducting the exploration of the trends in scientific and innovation fields (Wieczorek et al., 2021; Adam & Kogler, 2025; Kang 2023; Blei & Lafferty, 2006; Lu et al, 2022). By extracting latent themes from large text corpora, topic models enable researchers to identify dominant research areas, trace their growth trajectories, and detect shifts in scientific attention (Blei et al., 2003; Griffiths & Steyvers, 2004; Wieczorek et al., 2021). However, topics differ substantially in their maturity, scale, growth rate, and degree of novelty (Shiu et al., 2024). Treating all topics as equivalent risks obscuring important distinctions between established and consolidating fields and nascent, exploratory research directions (Small et al., 2014; Rotolo et al., 2015).

To address this limitation, recent scientometric research increasingly conceptualises relatively fast growth as the most frequently operationalised attribute of topic emergence (Potolo et al., 2023; Abercrombie et al., 2012). Building on this premise, scholars have employed three main approaches to identify emerging research topics: (i) detecting rapid growth in publication activity within predefined categories or controlled vocabularies; (ii) data-mining approaches that create relational structures from bibliographic data through co-occurrence clustering (e.g. co-author, co-word, or co-citation networks) and identify emergence within these structures; and (iii) hybrid approaches that combine growth-based indicators with structural or semantic clustering (Cozzens et al., 2010; Small et al., 2014).

The identification of rapid growth within existing categories or vocabularies is exemplified by burst-detection methods, most notably the algorithm proposed by Kleinberg (2002), which detects sudden increases in the frequency of terms or events over time. This approach has been widely implemented and extended in scientometric studies to identify emerging research fronts, including applications by Chen (2006), Börner et al. (2010), and Ohniwa et al. (2010). A second line of work focuses on the detection of emerging topics through structural changes in co-citation networks. For example, Upham and Small (2010) and Boyack and Klavans (2014) used co-citation clustering to identify newly forming or rapidly evolving research areas. Chen (2006; Chen et al., 2009; Chen et al., 2012) combined co-citation analysis with burst detection to characterise emerging trends in specific scientific domains, while direct citation approaches with temporal or annual clustering were proposed by Blei and Lafferty (2006) and Shibata et al. (2008).

Small et al. (2014) further argue that studies such as Roche et al. (2010) and Schiebel et al. (2010) represent a hybrid tradition by explicitly combining growth dynamics in vocabularies with rigorous cluster analysis. More recently, Lu et al. (2022) extend this hybrid line of research by integrating semantic filtering based on knowledge elements with bibliometric growth indicators, thereby moving beyond purely citation- or vocabulary-based clustering approaches.

Within this broader methodological landscape, one prominent approach to emerging-topic identification is the distinction between weak and strong signals, which differentiates nascent, exploratory topics from more established and rapidly expanding ones based on their growth dynamics and visibility (Kölbl & Grottke, 2020; Park & Kim, 2021; Zhao et al., 2024). Though the concept of strong and weak signals originates from business domain (Ansoff, 1975; Holopainen & Toivonen, 2012) and was aimed to be used as counteracting to strategic planning (Pépin et al, 2017; Mühlroth & Grottke, 2018), it was adopted in scientometrics (Jiang et al., 2025).

Ebadi et al. (2021) elaborated WISDOM, a multi-layered framework for topic identification and the systematic classification of strong and weak signals. WISDOM integrates advanced text representations, clustering techniques, and signal-detection frameworks, thereby extending traditional bibliometric approaches. By combining topic modelling with topic-signal analysis, the framework enables the identification of emerging research themes before they become mainstream. This added granularity is particularly valuable in technology domains shaped by geopolitical considerations—such as security, defence, and autonomous systems—where early awareness of emerging research directions can inform strategic decision-making and research policy.

Within this framework, strong-signal topics represent well-established and widely visible research areas, whereas weak-signal topics capture early-stage, low-volume, and potentially high-novelty research that may signal future breakthroughs (Ebadi et al., 2021; Ma et al., 2024). Strong-signal topics are characterised by high publication volume and sustained visibility, while weak-signal topics typically exhibit lower absolute volume but relatively high growth rates. Weak signals thus function as early indicators of potential structural change, pointing to emerging opportunities and new technological trajectories before they become widely recognised (Saritas, 2011; Holopainen & Toivonen, 2012).

In addition to strong and weak signals, WISDOM distinguishes latent signals, characterised by low volume and low growth, which may represent dormant or not-yet-activated research directions, and non-strong but well-known (NSWK) signals, which exhibit relatively high volume but limited growth and reflect mature or saturated topics. By jointly considering topic volume and growth dynamics, WISDOM provides a nuanced classification of research topics into established, consolidating, emerging, and potentially disruptive categories. Ebadi et al derive signals through the construction of Topic Emergence Maps (TEMs) and average calculation of topic propotion and growth rate.

Small et al. (2014) argue that there is always a reason that triggered the topic emergence. They distinguinsged between discovery in science, technological innovation and exogenous events to emergened topics that they identified.

## Methodology

### Data and methods

The data for this study were extracted from the OpenAlex 2026_01 snapshot in the MULTIOBS project and accessed via Google BigQuery. An initial exploratory search was conducted directly in OpenAlex to identify relevant concepts, topics, and terminology associated with drone research. Based on this exploration, a final query was executed in BigQuery with a comprehensive set of search terms, applying a keyword-based search to titles, abstracts, and author keywords. The query combined commonly used drone-related terminology, including "drone", "UAV", "UAS", "RPAS", "quadrotor", "multirotor", "aerial robotic(s)", "aerial robotics", "flying robot", "autonomous aerial system(s)", "low altitude platform" to capture the broader field of unmanned aerial systems research. The search was restricted to English-language journal articles published between 2020 and 2025 and assigned to the fields of engineering, energy, earth and planetary sciences, computer science, physics and astronomy, and materials science. Drone research is operationalised in a broad sense, encompassing UAVs and closely related autonomous aerial systems, including application domains and enabling infrastructures. The retrieval strategy prioritised broad recall in order to capture the heterogeneous and interdisciplinary nature of drone-related research. Consequently, the corpus may include a limited number of publications originating from adjacent technological domains such as robotics, sensing systems, precision agriculture, and autonomous systems.

The final dataset included 65,036 English-language articles. We included only articles with English-language metadata, as a preliminary application of BERTopic to a multilingual dataset resulted in substantial clustering driven by language rather than a topical structure. Metadata extracted for each article included the identifier, title, author keywords, abstract, author IDs, and institutional and country affiliations. Data processing, topic modelling, and visualisation were conducted using Python and R.

To identify strong- and weak-signal research topics, this study follows the methodology proposed by Ebadi et al. (2025), which combines transformer-based text embeddings with density-based clustering and topic salience measures. Textual content from titles, abstracts, and keywords was preprocessed using a customised stopword list that included both general English stopwords and domain-specific terms (e.g., "drone", "UAV", "system", "method") to reduce semantic noise. A count-based vectorizer with unigrams and bigrams was applied, retaining only terms appearing in at least three documents to retain low-frequency but potentially meaningful technical terms.

Document embeddings were generated using the SentenceTransformer model *paraphrase-multilingual-MiniLM-L12-v2*, which produces semantically meaningful sentence-level representations. This model was chosen to mitigate the risk of language-driven clustering, as, despite filtering for English-language metadata in OpenAlex, some abstracts may still include non-English content or multilingual elements. Dimensionality reduction was performed using Uniform Manifold Approximation and Projection (UMAP) (McInnes et al., 2018) with 30 nearest neighbours (n_neighbors = 30), five output dimensions (n_components = 5), cosine distance metric, and a minimum distance parameter of 0.0 (min_dist = 0.0) to preserve local semantic structure while enabling compact cluster formation. Topic clustering was subsequently conducted using the HDBSCAN algorithm with a minimum cluster size of 15 documents (min_cluster_size = 15) and a minimum sample parameter of 3 (min_samples = 3). The parameter prediction_data = TRUE was enabled to support soft clustering and outlier assignment. This BERTopic configuration was selected to balance topic specificity and cluster stability in a semantically heterogeneous and interdisciplinary corpus.

In total, the BERTopic model generated 468 topics. The complete list of topics is provided in the Online Appendix. Some topics were associated with broader governance and application domains rather than UAV-core engineering alone. This reflects both the interdisciplinary nature of contemporary UAV research and the retrieval strategy employed in OpenAlex, which used broad UAV-related conceptual phrases to capture the wider autonomous aerial systems ecosystem, including enabling technologies, application areas, and governance-related dimensions linked to drone research.

A large -1 Topic labelled “control unmanned aerial time algorithm” comprised approximately 47% (30,588 articles) of the corpus. Additional clustering attempts were conducted by reducing the minimum topic size parameter from 15 to 10 and subsequently to 8 documents; however, these adjustments did not substantially reduce the size of -1Topic. Inspection of representative titles within the residual cluster of -1 topic revealed that it captures a broad methodological and engineering-oriented layer of drone research spanning autonomous control, reinforcement learning, navigation systems, cybersecurity, aerodynamics, computer vision, and communication systems. Although these publications share a common UAV-related technical vocabulary, they do not necessarily form sharply bounded thematic communities, which likely contributed to their assignment to the residual outlier cluster by the density-based HDBSCAN algorithm. Consequently, the original model results were retained for the subsequent analyses.

For each topic, descriptive labels were drawn from the top-ranked words identified by BERTopic (Grootendorst, 2022). In addition to topic labels, the top 10 words and example documents were extracted for each topic to support qualitative validation and transparency. Topic labels were applied solely for descriptive interpretation and did not influence signal classification.

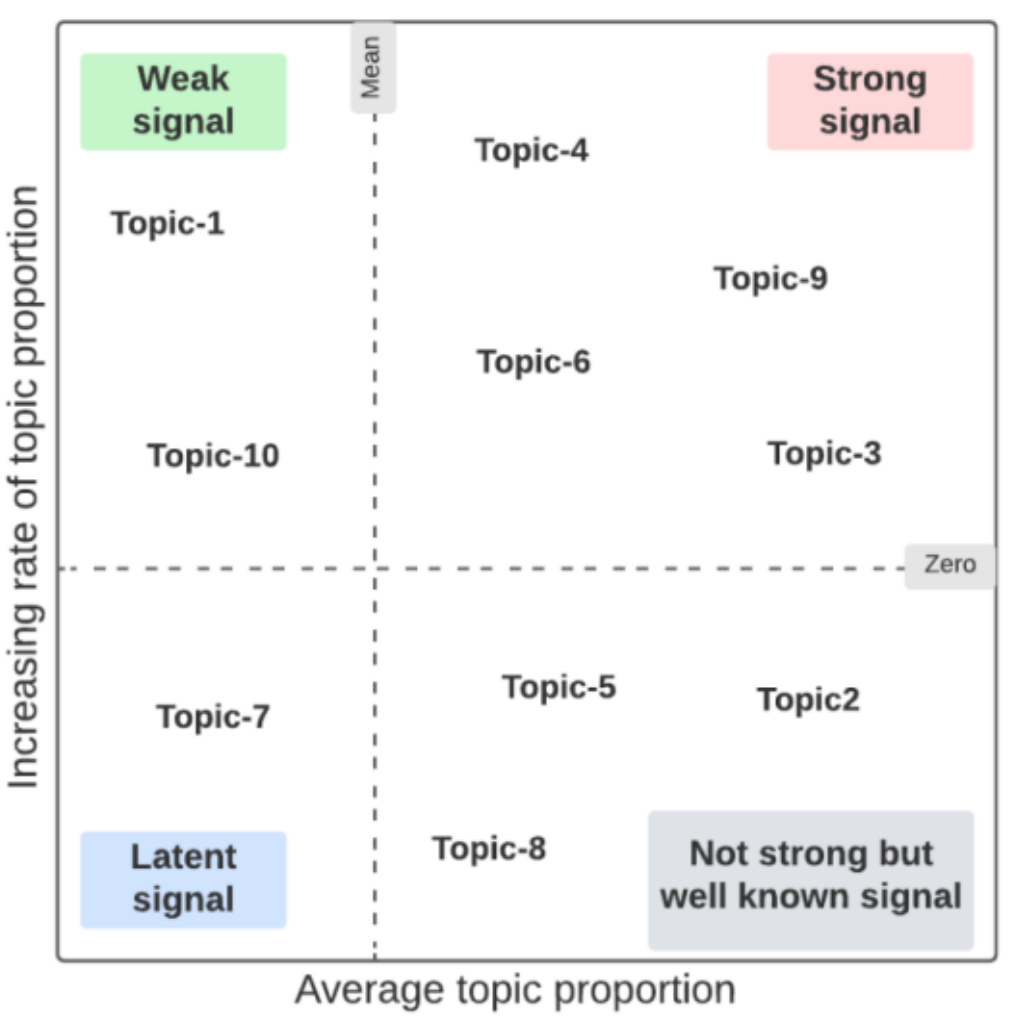

**Fig. 1** Topic emergence map (TEM), detecting future signals. The TEM is divided into four quadrants based on the mean of values on the x-axis and a y-axis value of zero (retrieved from Edabi et al., 2025).

The WISDOM framework applied in this study (Ebadi et al., 2025) classifies topics based on average topic proportion and growth rate. The average topic proportion measures a topic’s relative presence across all documents in a defined period (2020-2025), while the growth rate captures the relative increase in topic prominence during this period. Using the mean topic proportion and a growth rate of zero as thresholds, topics are classified into four categories: strong signals (proportion above the mean, positive growth), weak signals (proportion below the mean, high growth), latent signals (proportion below the mean, low growth), and non-strong but well-known signals (NSWK) (proportion above the mean, low growth). Strong-signal topics represent well-established and consolidated areas of drone research, whereas weak-signal topics capture emerging, fragmented, or exploratory thematic areas with lower visibility. Of the 34,448 articles, 11,827 (18% of the dataset) were assigned to 57 strong-signal topics and 4,815 articles (7.4% of the dataset) were assigned to 154 weak-signal topics. The remaining 17,806 articles (27.4% of the dataset) were classified either as latent or NSWK.

***Measuring fractional authorship contribution and citation impact across countries***

Fractional authorship contribution across countries was calculated based on the country affiliations of all authors for each article and subsequently aggregated into focal categories: China, the United States, the EU-14, the EU-13, Ukraine, Russia, and "Other". The study distinguishes between the EU-14: long-standing Western and Southern European member states prior to the 2004 enlargement (Austria, Belgium, Denmark, Finland, France, Germany, Greece, Ireland, Italy, Luxembourg, Netherlands, Portugal, Spain and Sweden); the EU-13: the Central and Eastern European countries that joined the EU in 2004 or later (Bulgaria, Croatia, Cyprus, Czechia, Estonia, Hungary, Latvia, Lithuania, Malta, Poland, Romania, Slovakia, and Slovenia).

Field-Normalised Citation Impact (FNCI) was calculated for the full dataset of English-language drone-related articles indexed in OpenAlex MULTIOBS 2026. For citation analysis, only articles published between 2020 and 2024 were included, while articles from 2025 were excluded due to a relatively small citation window. First, the average citation count was computed separately for each publication year (2020–2024). Second, for each article in the dataset, the observed citation count was divided by the corresponding yearly average, yielding a time-normalised citation impact measure.

To examine systematic differences in citation impact, two linear regression models were estimated using the log-transformed FNCI as the dependent variable (log(FNCI + 1)). The log transformation was applied to reduce the influence of extreme citation values and to account for the highly skewed distribution of citation counts, following established bibliometric practice (Thelwall & Wilson, 2014). Independent variables of Model 1 included country, publication year and their interaction. Independent variables in Model 2 comprised country, authorship type, signal type (strong vs. weak), and publication year, along with their interactions.

Citation exchange between countries was analysed as a directed weighted network based on citation links between publications. Citation flows were aggregated into a bilateral citation matrix Cij, where Cij represents citations from country i to country j, and Cji represents citations in the opposite direction. To examine directional asymmetries in knowledge exchange, a normalised Knowledge Exchange Balance (KEB) indicator was constructed as:

$$B_{ij} = \frac{C_{ij} - C_{ji}}{C_{ij} + C_{ji}}$$

where positive values indicate that country i cites country j more than it is cited by j, while negative values indicate the opposite. The indicator ranges from −1 to +1, with values closer to zero reflecting a more balanced bilateral citation exchange. Unlike aggregate knowledge trade indicators based on ratios of references provided to references received, the present approach focuses on dyadic citation asymmetries between specific country pairs, allowing direct comparison of bilateral knowledge exchange structures.

Citation links were fractionalised at both the citing and cited sides to reduce inflation associated with multi-country co-authorship. Records with missing affiliation information were retained as a separate "Unknown" category to preserve the overall citation structure and avoid bias introduced by listwise deletion.

## Results

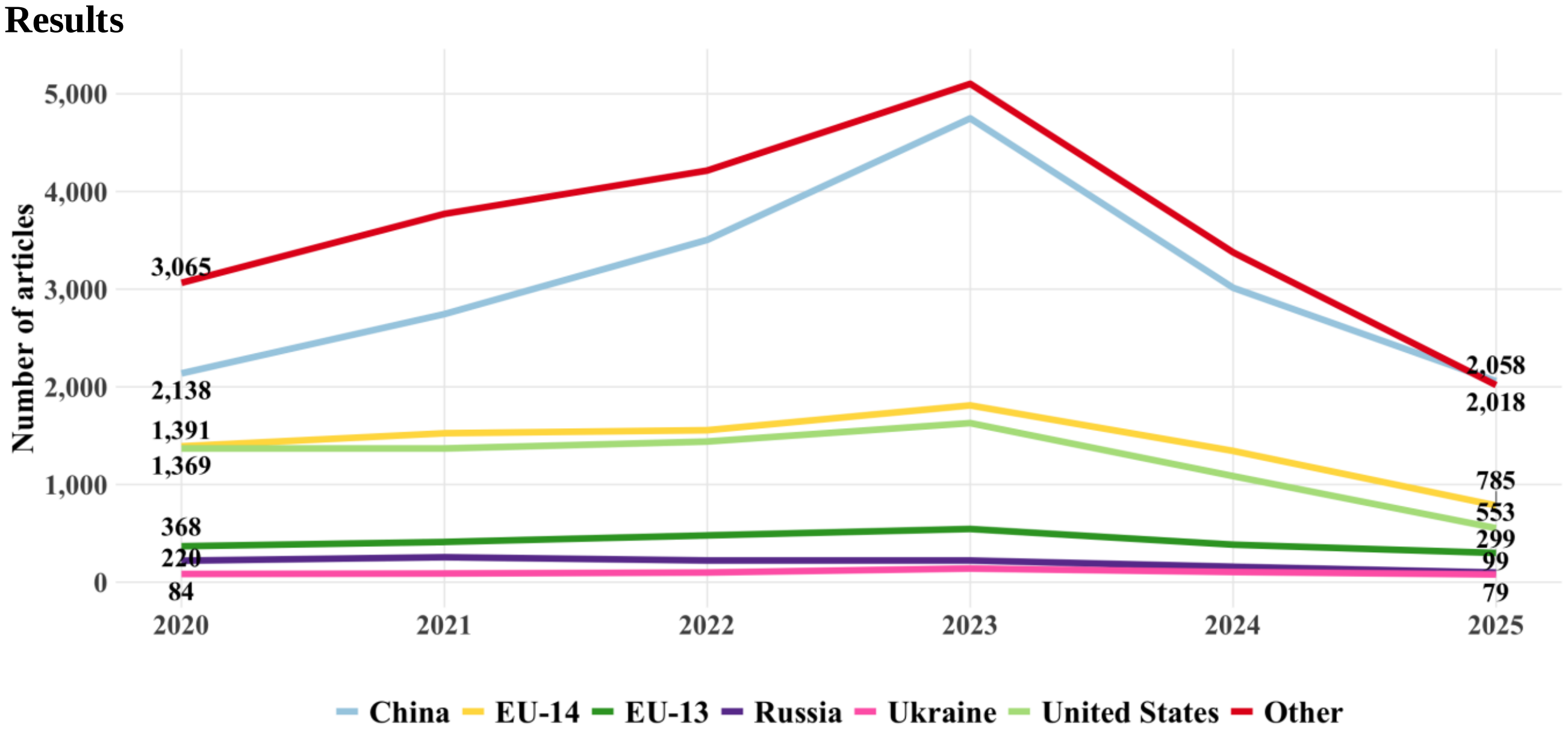

**Fig. 2** Number of articles by country

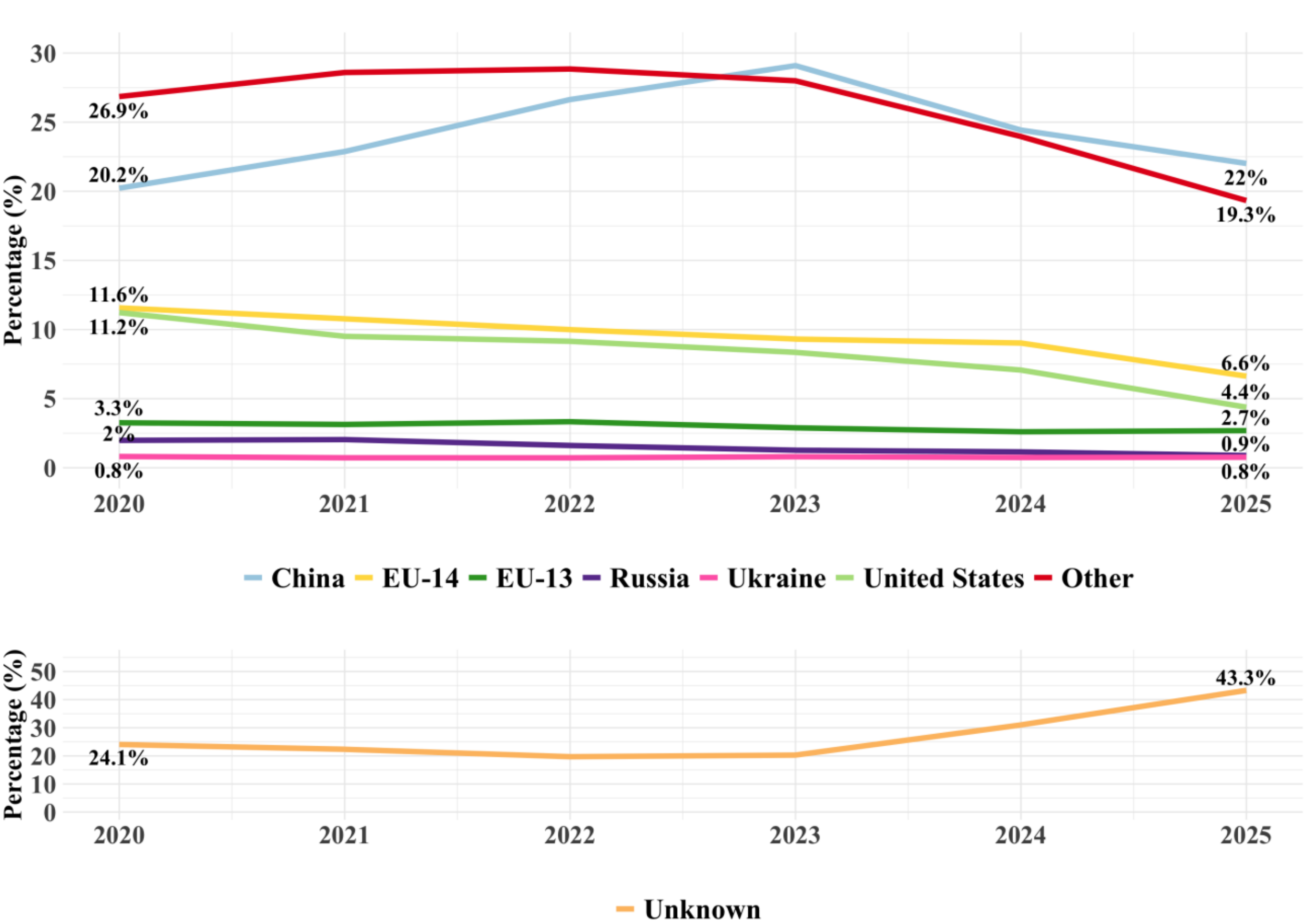

**Fig. 3** Fractional authorship contribution across countries

Figure 2 shows the dynamics of the number of articles across countries. Publication activity increased substantially between 2020 and 2023 for all countries, followed by a decline in 2024–2025. The strongest growth was observed for China and the heterogeneous "Other" category, both of which substantially outpaced the United States and EU-14 countries in absolute publication volume. In contrast, the United States, EU-14, and EU-13 countries exhibited more moderate growth trajectories, while Russia and Ukraine remained comparatively marginal contributors throughout the period.

Figure 3 presents the fractional authorship contribution across countries, highlighting the dominance of China, whose share reached  22% in 2025. In contrast, the fractional authoship contribution from the EU-14 countries and the United States decreased over time. The growing share of articles with 'Unknown' affiliations reflects incomplete institutional disambiguation in OpenAlex, particularly in recent years.

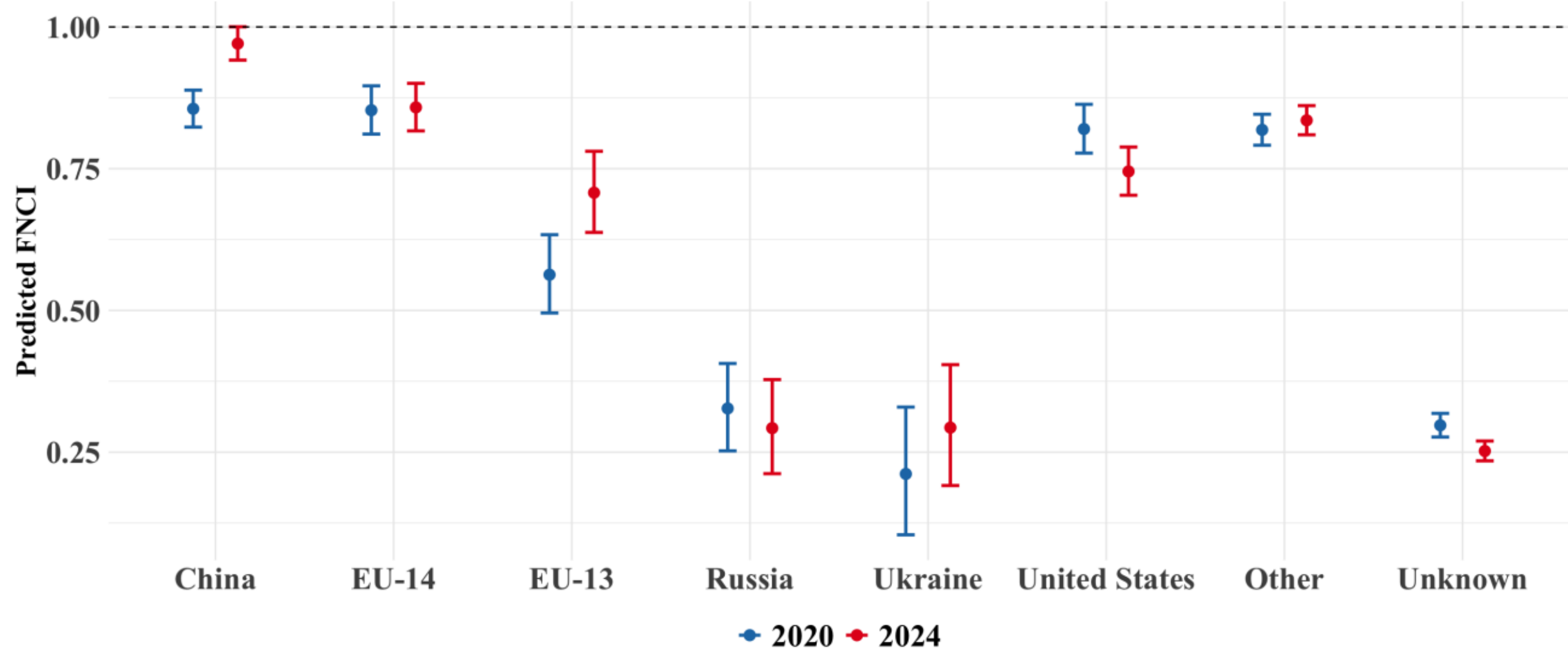

**Fig. 4** Predicted FNCI across countries

Model 1, a linear regression model predicting log-transformed field-normalised citation impact (log(FNCI + 1)), was fitted to explore the effects of geographic entity, publication year and their interaction on citation impact. The logarithmic transformation was applied to reduce the influence of highly skewed citation distributions and extreme values, which are characteristic of bibliometric data. The model explains approximately 7.7% of the variance (adjusted $R^2 = 0.077$), which is consistent with typical findings in bibliometric analyses, given the high variability in citations. Table 1 in Appendix presents the Model 1 summary, and Table 2 presents the predicted FNCI.

The predicted FNCI values derived from Model 1 (Figure 4) reveal substantial cross-national differences in citation impact. China exhibited the strongest increase between 2020 and 2024, with predicted values rising from 0.86 to 0.97. EU-14 countries remained comparatively stable, with predicted FNCI at 0.85 and 0.86. In contrast, the United States showed a moderate decline in predicted FNCI over time. EU-13 countries showed an increase, though the values remained rather moderate. Russia and Ukraine consistently displayed substantially lower predicted FNCI, while the heterogeneous "Other" category maintained relatively high and stable values.

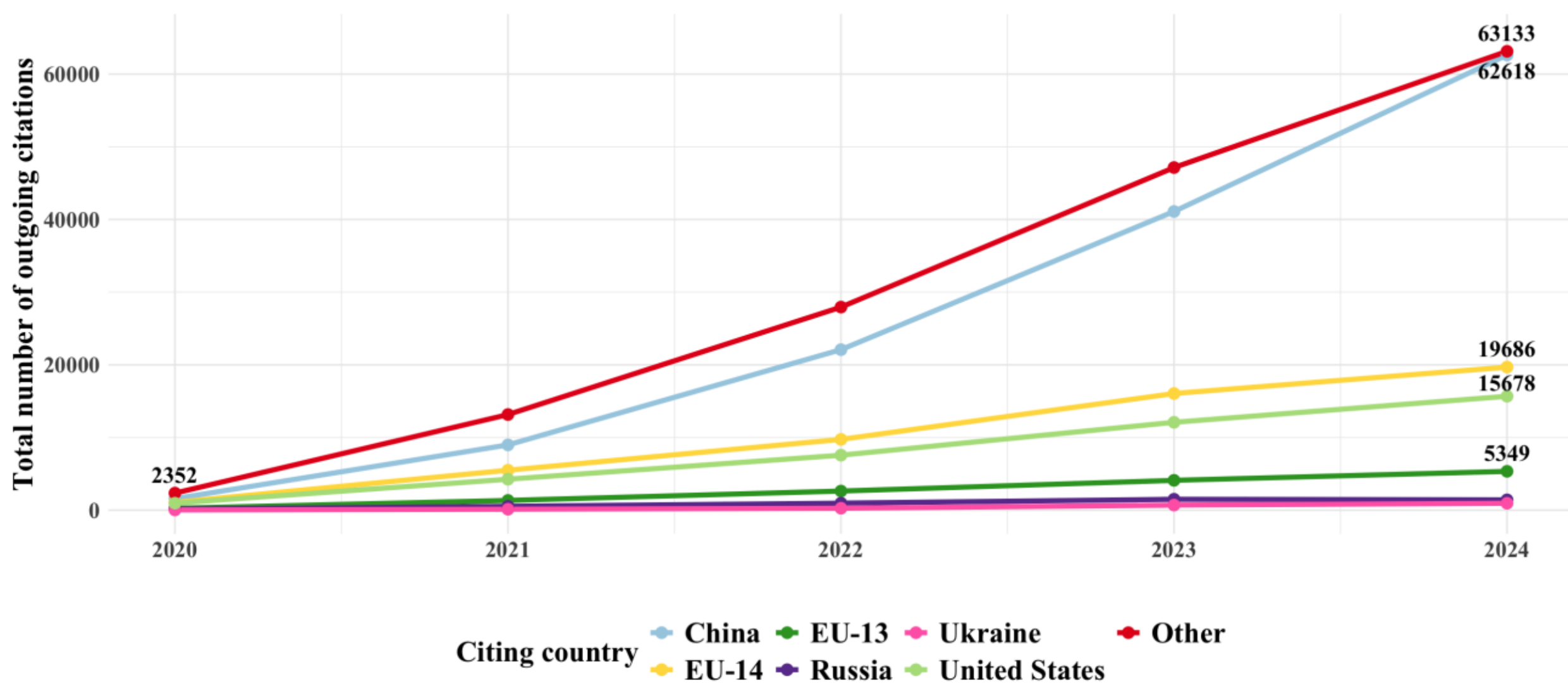

**Fig. 5** Citations across citing countries

Fig 5 shows that China produced the largest number of citations within the global drone-related citation system between 2020 and 2024. Although citation activity increased across all focal country groups, the substantially steeper expansion of citations from China-affiliated publications indicates a rapid increase in the scale and centrality of China-linked knowledge circulation within the field. At the same time, the persistently large share of citations originating from the "Other" category suggests that drone-related scientific exchange remains globally distributed rather than fully regionalised.

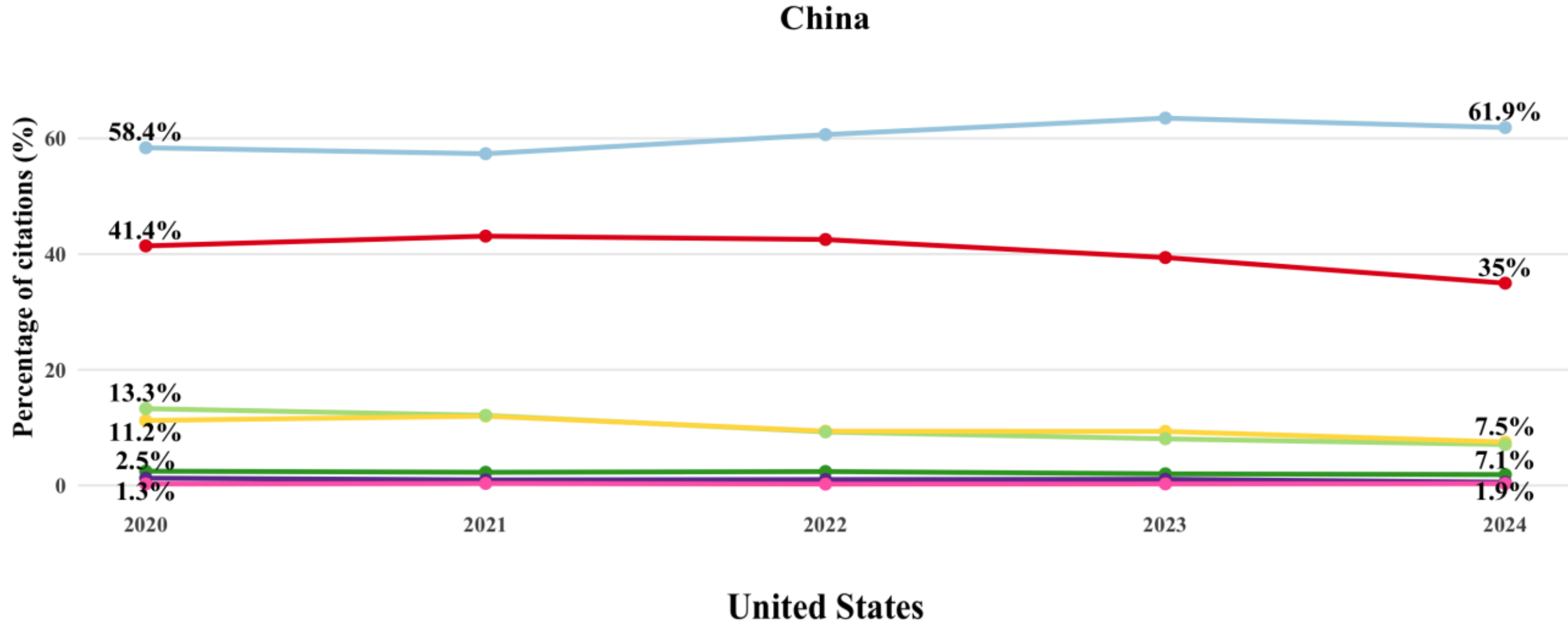

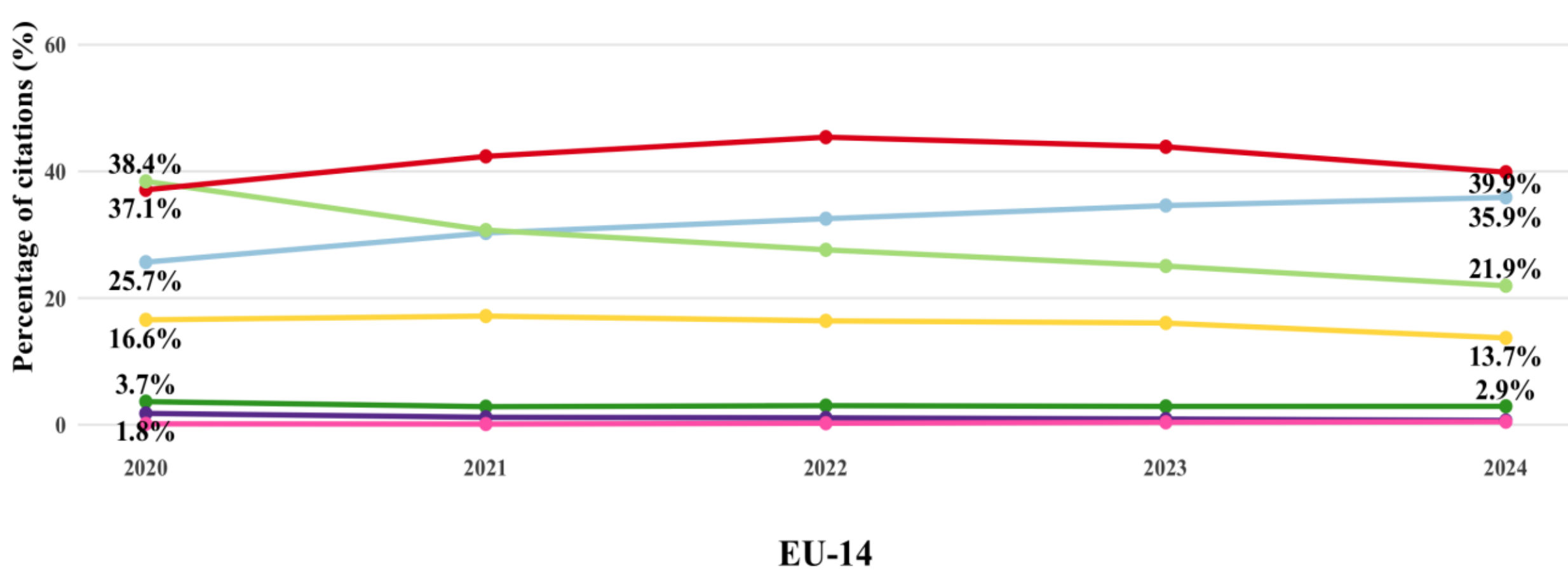

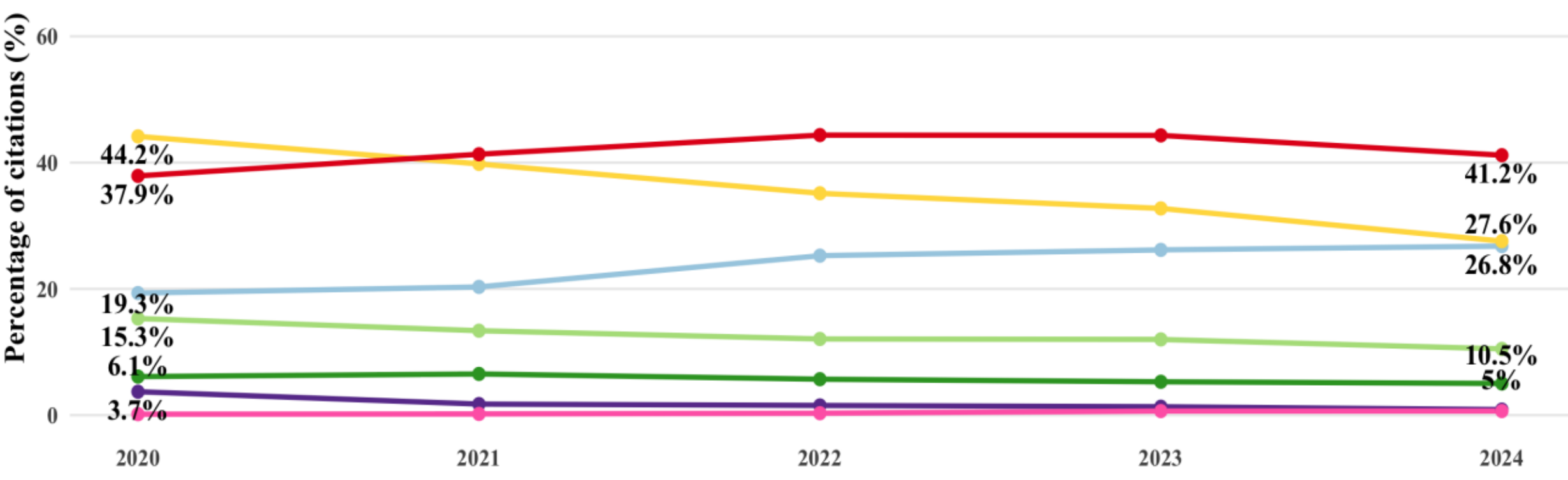

EU-13

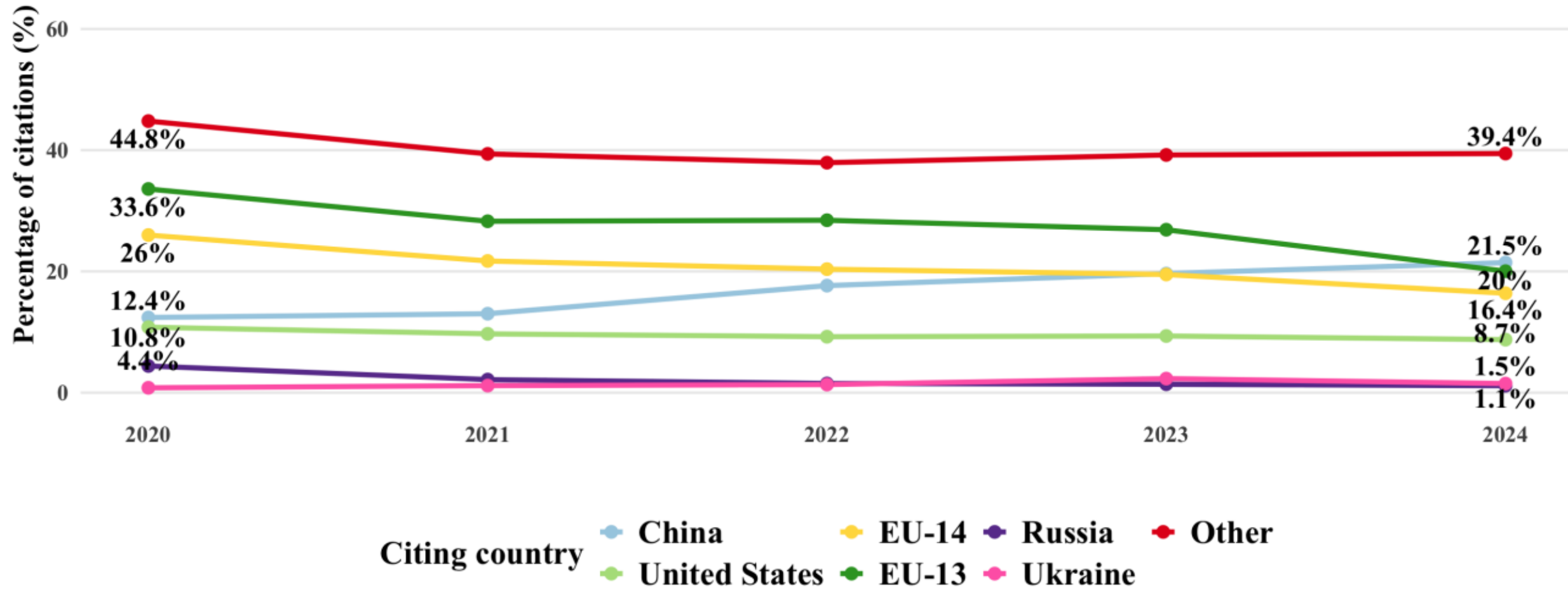

**Fig. 6** Citation inflows to China, United States, EU-14 and EU-13 by citing country

Between 2020 and 2024, citations originating from China remained the dominant source of citations to China-affiliated publications and increased both in absolute terms and relative share, indicating growing domestic circulation within the Chinese drone research field. At the same time, citations originating from the United States, the EU-14, and the broader "Other" category also increased in number, suggesting that stronger domestic citation concentration coexisted with continued international visibility, although their overall shares decreased.

In contrast, citations to United States-, EU-14-, and EU-13-affiliated publications remained more internationally distributed. Although domestic citation volumes increased over time, their relative shares generally declined, indicating increasing diversification of citation sources. Simultaneously, the share of citations originating from China increased across all three country groups, particularly for the United States, reflecting the growing prominence of China-affiliated research within global drone-related citation networks.

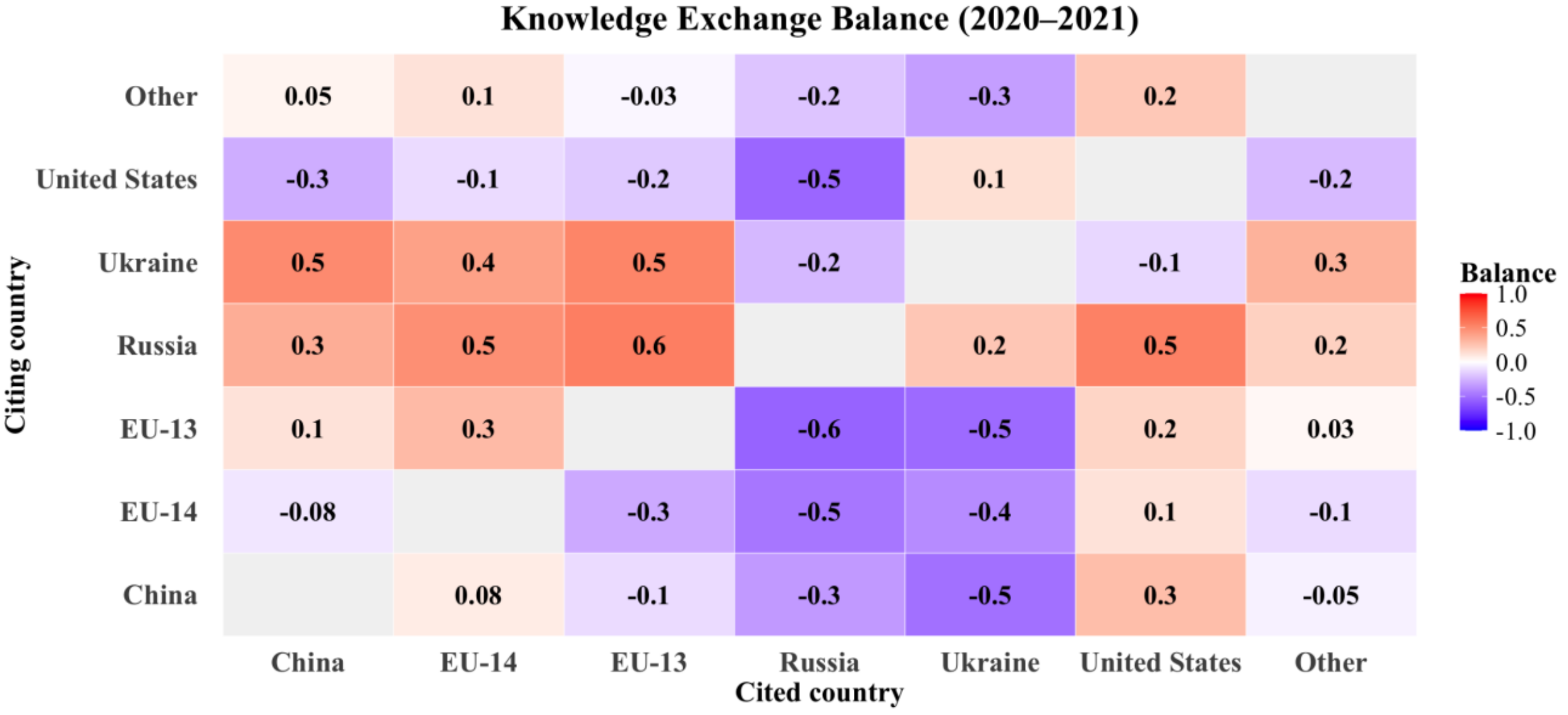

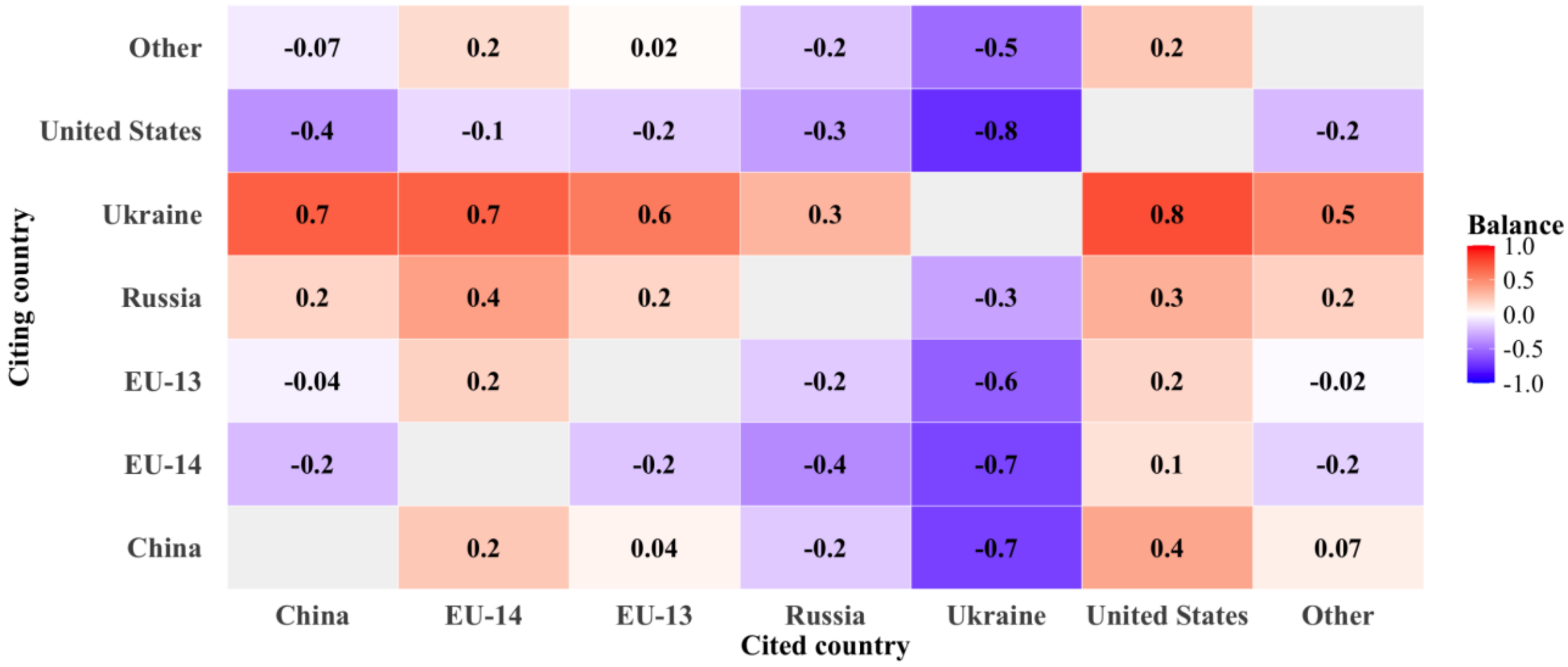

**Fig. 7** Knowledge exchange balance

Fig. 8 illustrates the dynamics of the Knowledge Exchange Balance (KEB) between countries across two periods. Positive values indicate that a country cites another country more frequently than it is cited in return.

The EU-14 and the United States showed mostly positive KBE, with the EU-14 consistently citing the United States more often than the reverse across both periods. This pattern remained stable over time. China exhibited predominantly positive KEB values toward both the EU-14 and the United States, indicating a stronger outbound citation orientation toward these regions, which further increased across periods.

The EU-13 reduced a negative balance with the EU-14 across periods and got a marginal positive KEB from China. In contrast, Russia and Ukraine are characterised by predominantly negative balances throughout both periods.

## Strong- and weak-signals topics

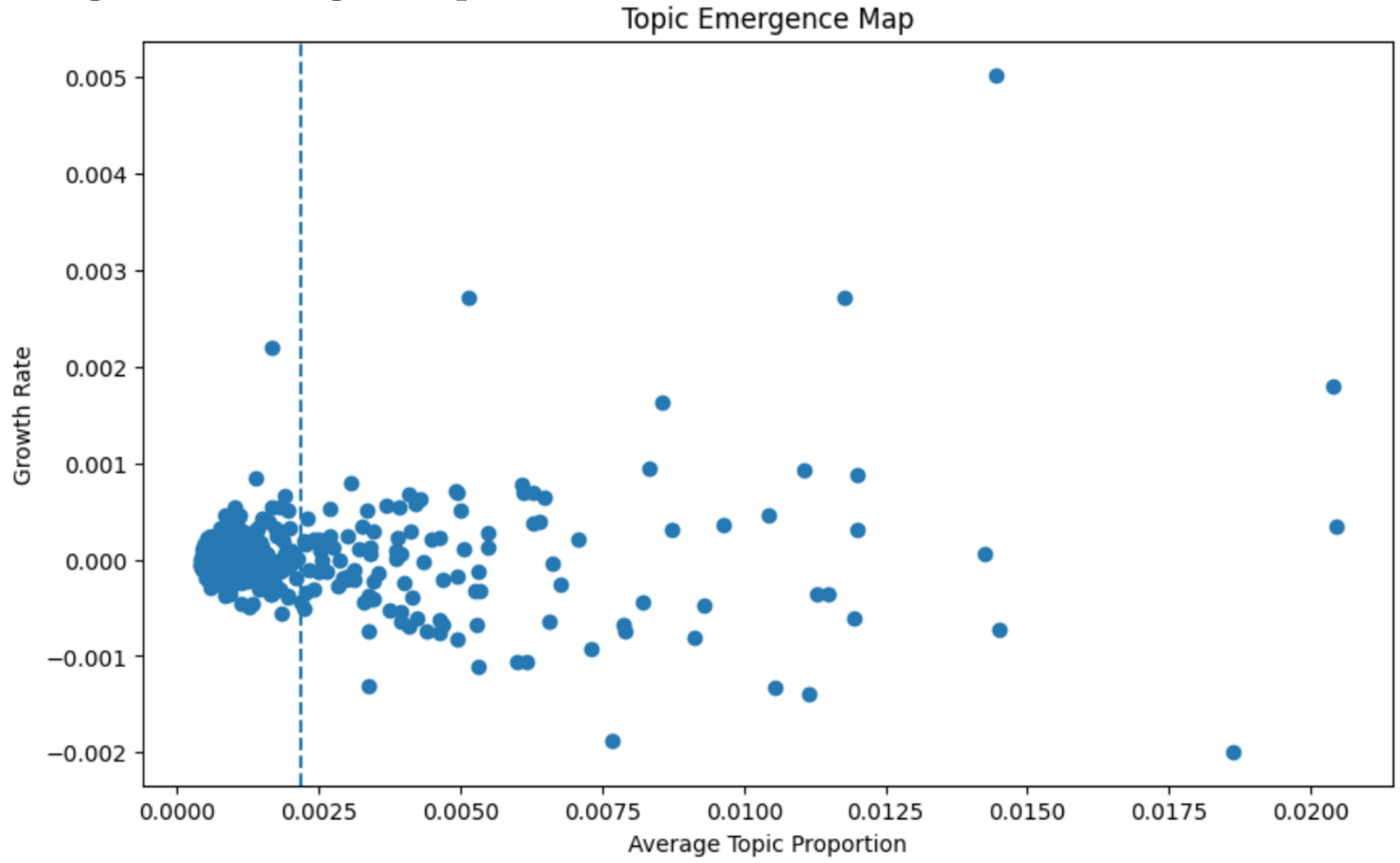

**Fig. 9** Topic emergence map

Figure 8 presents the  topic emergence map, which reveals a highly concentrated topic structure, with the majority of topics clustered near the origin, indicating relatively low average topic proportions and limited temporal variation over the 2020–2025 period. Only a small number of topics exhibit strong positive growth rates, suggesting a limited set of rapidly emerging research directions. Several topics located in the upper-right quadrant can be interpreted as strong signals, characterised by both high prevalence and increasing temporal prominence. In contrast, topics positioned in the bottom-right quadrant represent established but declining themes (NSWK signals), while topics in the upper-left quadrant indicate weak but potentially emerging signals. Overall, the distribution suggests a research landscape dominated by stable niche topics, accompanied by a smaller subset of expanding thematic areas.

Topics with the highest growth rate include such strong-signal topics as "Object detection visdrone improved yolov map" (growth rate = 0.005; top words: object detection, visdrone, improved yolov, map, small, backbone, dataset, head, pyramid, fusion) and "Armed forces battlefield conflicts" (growth rate = 0.0027; top words: armed, forces, battlefield, conflicts, tactical, operations, strikes, federation, Ukraine war, invasion), as well as weak-signal topics such as "Object detection visdrone convolution feature fusion" (growth rate = 0.0007; top words: object detection, visdrone, convolution, feature fusion, module, linguistics, datasets, small objects, scale feature, backbone) and "Geo localization cross matching satellite" (growth rate = 0.0005; top words: geo localization, cross, matching, satellite, feature, university, view image, transformer based, datasets, existing methods).

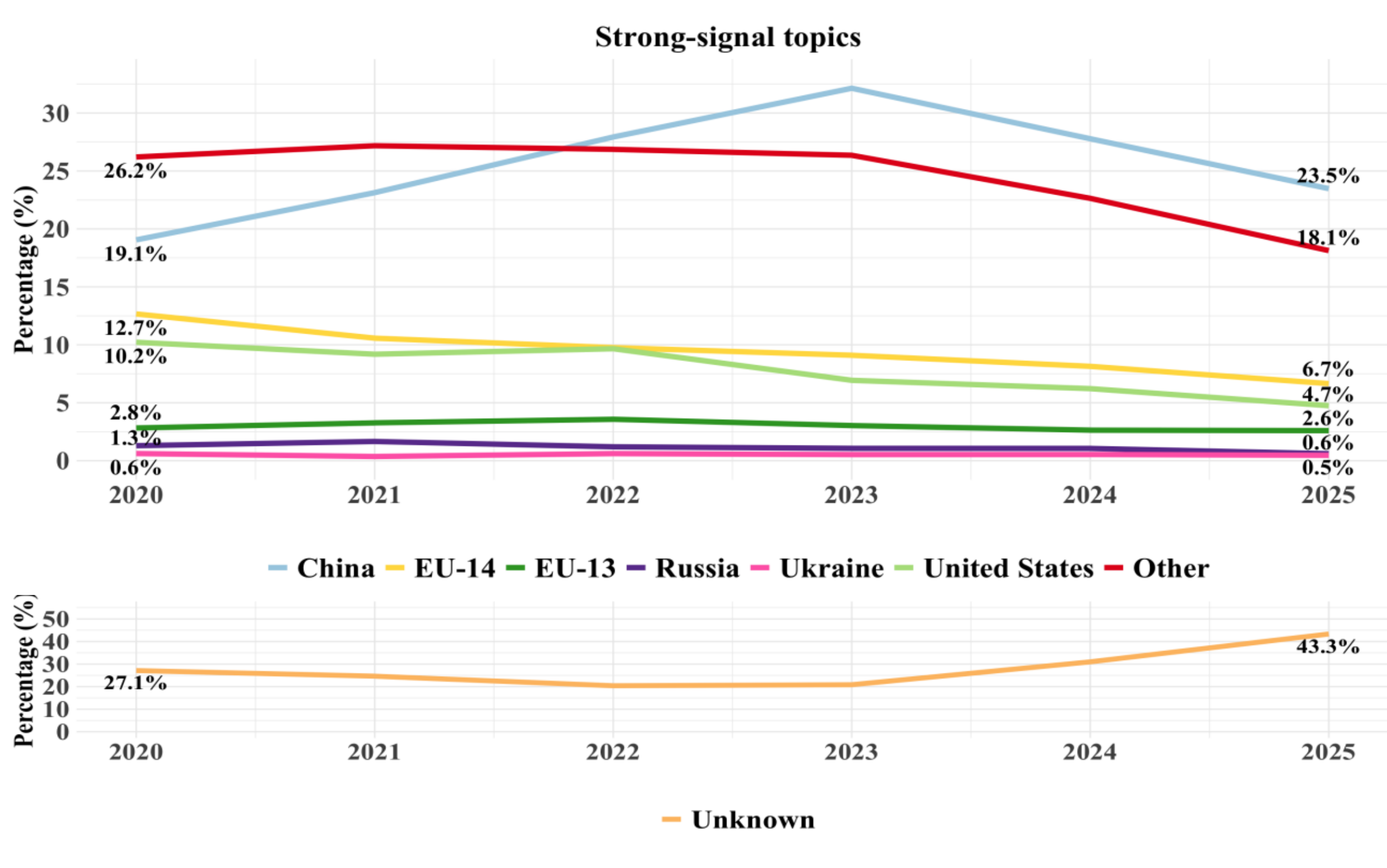

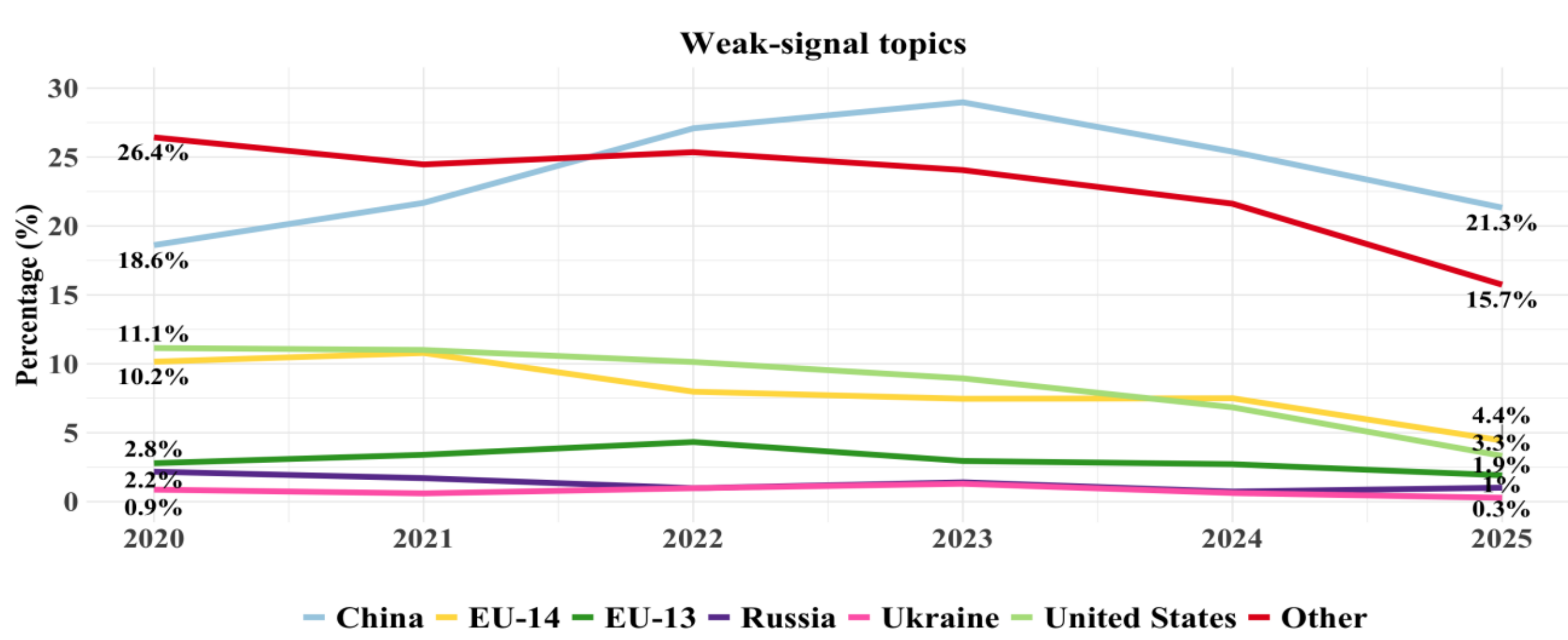

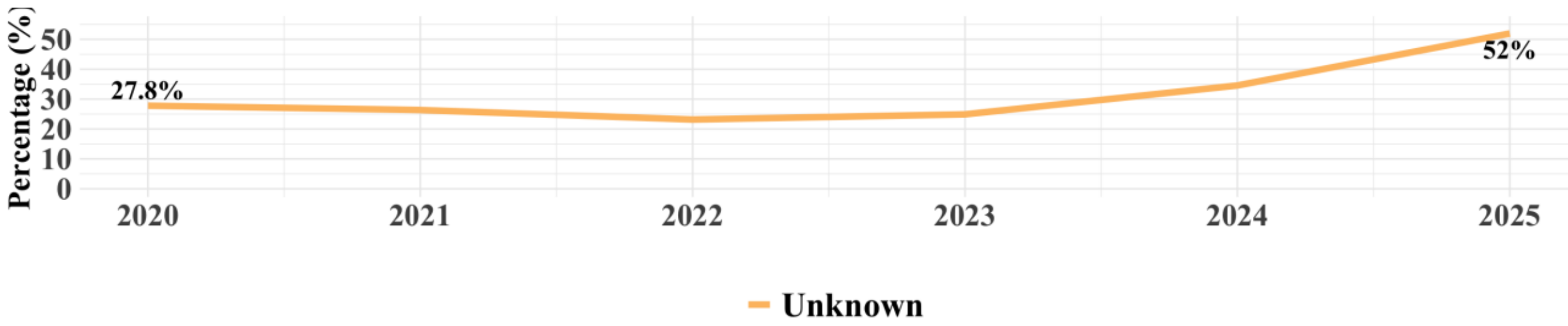

**Fig. 8** Fractional authorship contribution across countries in strong- and weak-signal topics

Figure 7 illustrates the temporal dynamics of fractional authorship contribution across countries in strong- and weak-signal topics. In both cases, China exhibited a substantial increase and dominance, with a higher share in strong-signal topics (23.5) than weak-signal ones (21.3%) in 2025. By contrast, the share attributed to "Other" countries decreased over the observed period. The United States and EU-14 displayed moderate but gradually declining participation, while Russia and Ukraine maintained comparatively marginal shares throughout the period. Overall, the fractional authorship contributions across countries across in two topic types were rather similar. However, the proportion of articles with unidentified affiliations ("Unknown") increased more noticeably by 2025 within weak-signal topics, suggesting higher metadata incompleteness.

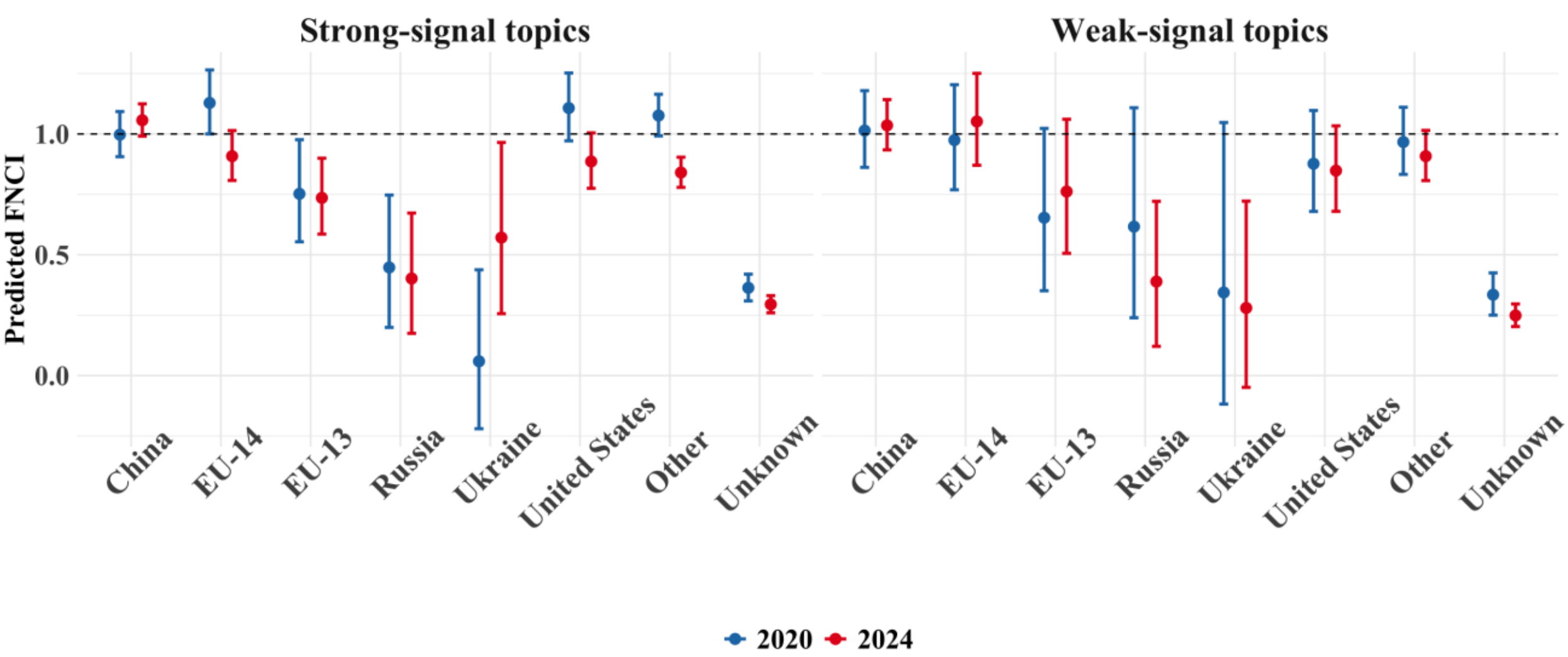

**Fig. 10** Predicted FNCI by country across topic types

Model 2, a linear regression model predicting log-transformed normalised citation impact (log(FNCI + 1)), was fitted to explore the effects of country, topic type, publication year and their interactions on FNCI. The log-transformation was applied to account for the strong right-skewness typical of citation distributions and to reduce the influence of extreme values. The model explains approximately 8.3% of the variance (adjusted $R^2 = 0.083$), which is consistent with typical findings in bibliometric analyses, given the high variability in citation counts.

Table 3 in Appendix presents the Model 2 summary, and Table 4 presents the predicted FNCI. Figure 11 illustrates the predicted FNCI values derived from Model 2. From 2020 to 2024, China exhibited growth in predicted FNCI across both strong-signal (from 0.997 to 1.057) and weak-signal topics (from 1.014 to 1.036). In contrast, the United States showed a decline in both topic types, with predicted FNCI values falling below the world average benchmark of 1.0 by 2024. The EU-14 showed divergent trends across topic types: predicted FNCI increased from 0.974 to 1.052 in weak-signal topics but declined from 1.128 to 0.908 in strong-signal topics. Consequently, the EU-14 slightly lost its citation impact advantage over China in strong-signal topics, while retaining a marginal advantage in weak-signal topics. This pattern is also reflected in the upper bounds of the 95% confidence intervals. This finding suggests that the EU-14 retains a relatively strong role in emerging and potentially innovative research topics. The EU-13 countries showed intermediate levels of citation impact levels, but their performance remained consistently below the global average in both topic types.

Russia, Ukraine, and the “Unknown” category displayed substantially lower predicted citation impact across both topic categories. Notably, Ukraine showed a marked increase in predicted FNCI in strong-signal topics between 2020 and 2024, although confidence intervals remained comparatively wide, indicating greater estimation uncertainty. The “Other” category demonstrated comparatively high predicted FNCI values in both topic types.

Overall, weak-signal topics exhibited wider confidence intervals, indicating greater variability and uncertainty in citation performance.

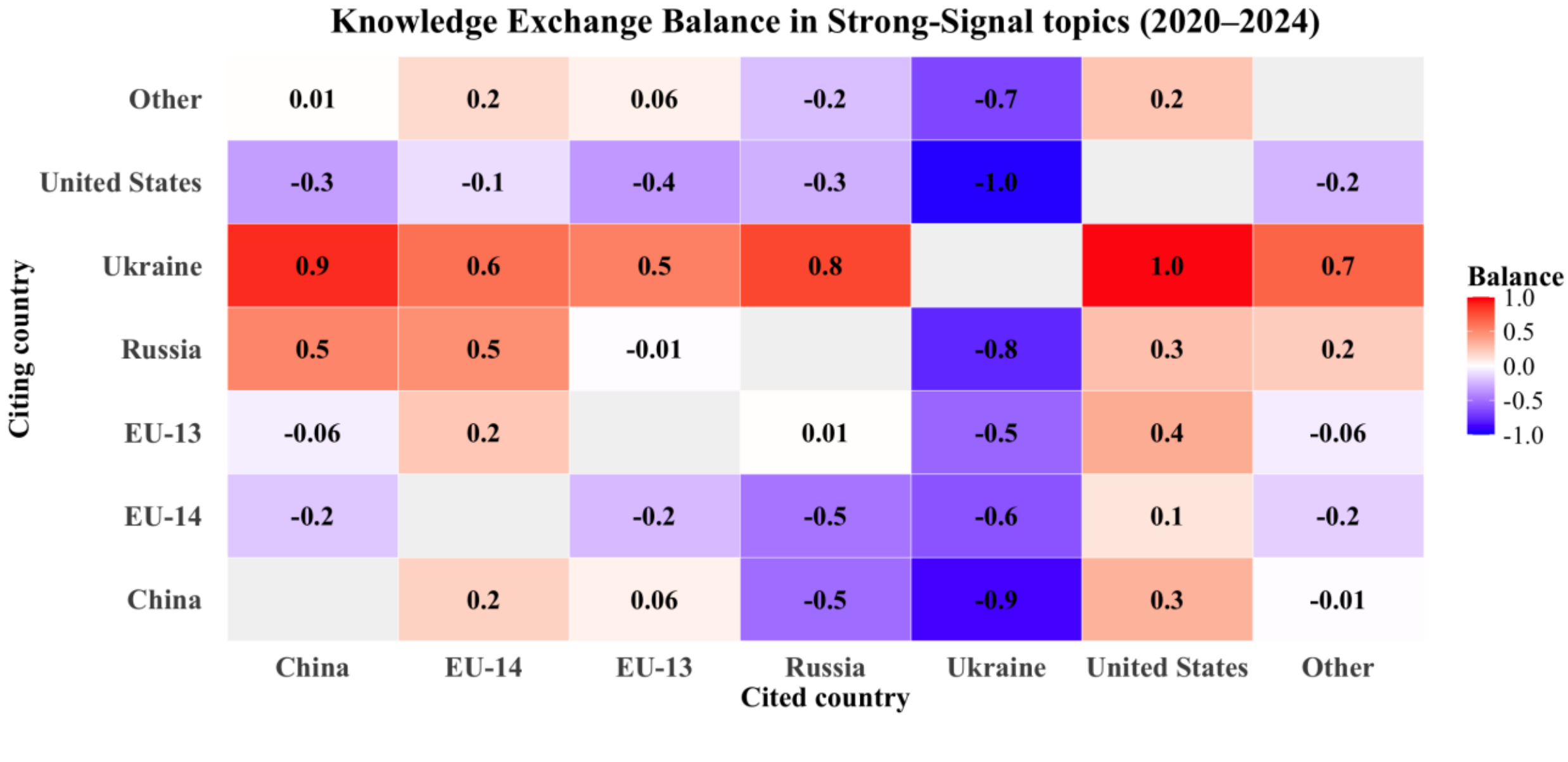

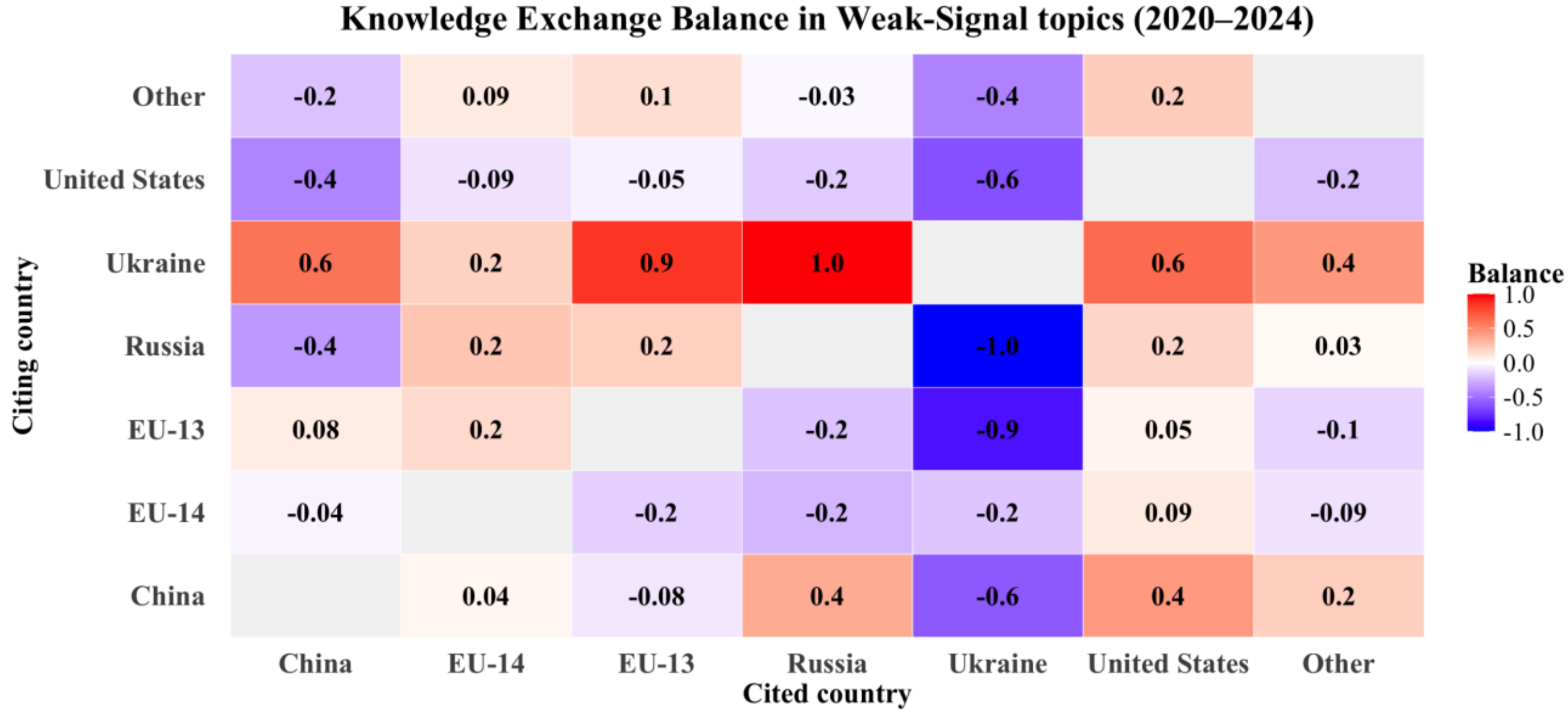

**Fig. 11** Knowledge Exchange Balance (KEB) across strong- and weak-signal topics

Fig. 11 illustrates the Knowledge Exchange Balance (KEB) between countries across strong- and weak-signal topics. In both topic categories, the EU-14 and the United States exhibited largely balanced and positive bilateral citation exchange, with the EU-14 consistently citing the United States slightly more frequently than the reverse. Notably, the magnitude of this imbalance remained similar across both strong- and weak-signal topics, suggesting relatively stable transatlantic citation relationships across established and emerging research areas.

China exhibited negative KEB with both the EU-14 and the United States, particularly with the latter, indicating that China-affiliated publications cited these research systems more frequently than they were cited in return. This asymmetry was more pronounced in weak-signal topics, where citations from China to the United States exceeded those directed toward the EU-14, suggesting a stronger orientation toward US-linked research within emerging thematic domains.

The EU-13 exhibited negative KEB with the EU-14, the United States, and China (weak-signal topics), but positive KEB with Russia and Ukraine, reflecting comparatively stronger regional citation exchange patterns. Russia and Ukraine generally functioned as net exporters of citations within the analysed citation relationships.

KEB dynamics were not explored temporally within topics due to the comparatively smaller number of citation links in weak-signal topics, which could produce unstable bilateral estimates and exaggerated year-to-year fluctuations.

**Conclusions**

The study findings revealed that drone-related science is characterised by growing geopolitical asymmetries in scientific production, citation concentration, and international knowledge exchange. In particular, China increasingly dominated scientific production, fractional authorship contribution, and domestic citation circulation. In contrast, the United States and EU countries maintained comparatively more internationally distributed citation structures. However, China-affiliated publications also became increasingly integrated within global citation networks, particularly through growing citation exchange with the United States and European countries.

The peak in publication output observed in 2023 coincided with the period following Russia's full-scale invasion of Ukraine, which may have accelerated global scientific and technological interest in drone-related systems and associated R&D activities. Despite the strategic importance of drones in the military operations of both Russia and Ukraine, the scientific publication output of these countries remained comparatively modest. This may reflect security-related restrictions, limited dissemination of sensitive research, or a divergence between actors engaged in operational drone development and those publishing research on drones.

Notably, interpretation of authorship and citation patterns is complicated by the high proportion of publications with unidentified affiliations, particularly within weak-signal topics, where the share of "Unknown" affiliations exceeded 50% by 2025. More broadly, the increasing proportion of unidentified affiliations in the overall dataset, reaching 43% in 2025, highlights persistent limitations in institutional disambiguation within OpenAlex. These findings underscore the importance of developing comprehensive Research Organisation Registry (ROR) infrastructures and improving institutional name standardisation and affiliation disambiguation practices to enhance attribution accuracy in large-scale open bibliometric databases.

Although China demonstrated a citation advantage, it has exhibited a strong tendency toward domestic citation concentration, with a substantial proportion of citations originating from China-affiliated publications. This suggests that China's citation advantage is partly driven by high internal domestic citation concentration rather than exclusively by global integration. This aligns with Qiu et al (2025). Moreover, China imported proportionally more knowledge from the EU-14/US than it exported back, with this asymmetry increasing over time. EU-13 countries demonstrated comparatively weaker publication and citation positions than the EU-14 and exhibited higher reliance on external knowledge flows from the EU-14 and the US. However, EU-13 shifted from exporter to importer of citations from China, though the extent of citations imported was rather modest.

Topic evolution analysis indicated a predominantly incremental rather than disruptive development pattern in drone research. The thematic landscape is characterised by a limited number of rapidly expanding topics alongside a large set of stable research areas, suggesting gradual rather than radical structural change in the field. Topic-level analysis further showed that drone research is characterised by substantial thematic heterogeneity and interdisciplinarity. The findings highlight that contemporary UAV research increasingly operates as a distributed technological ecosystem spanning both technical and application-oriented domains. Notably, EU-14 countries maintained the strongest citation impact in weak-signal topics, suggesting a more prominent role in shaping emerging research directions. At the same time, China-affiliated publications cited the United States more frequently than the EU-14 in both strong- and weak-signal topics, with this pattern being particularly pronounced in weak-signal areas. This suggests that, despite the citation influence of the EU-14 in emerging domains, the United States remains China's primary external knowledge source within the global drone research system.

Finally, the application of the WISDOM model revealed methodological limitations associated with the large number of publications assigned to the residual "-1Topic" outlier category. Attempts to further subdivide this category by reducing the minimum clustering threshold from 15 to 10 and subsequently to 8 publications did not produce sufficiently coherent subtopics. It highlights the need for further development of robust topic modelling approaches for heterogeneous and rapidly evolving research fields.

## Policy implications

The findings of this study have implications for research information systems, research evaluation, and science policy. First, research evaluation frameworks should account for the extent of domestic citation concentration. Second, the stronger performance of EU-14 countries in weak-signal (emerging) drone topics suggests that targeted support for early-stage and exploratory research may contribute to sustained scientific influence in emerging technological domains. Maintaining such performance may require continued investment in high-risk, high-uncertainty research portfolios.

## Declarations

Conflict of interest. The author claims no conflict of interest.

## Funding declaration

The author received no funding for this research.

## References

Abercrombie, R. K., Udoeyop, A. W., & Schlicher, B. G. (2012). A study of scientometric methods to identify emerging technologies via modeling of milestones. *Scientometrics*, *91*(2), 327-342.

Abramo, G., D'Angelo, C. A., & Di Costa, F. (2019). A nation's foreign and domestic professors: which have better research performance?(the Italian case). *Higher Education*, *77*, 917-930.

Adam, R., & Kogler, M. (2025, July). Bidirectional Topic Matching: Quantifying Thematic Intersections Between Climate Change and Climate Mitigation News Corpora Through Topic Modelling. In *Proceedings of the 2nd Workshop on Natural Language Processing Meets Climate Change (ClimateNLP 2025)* (pp. 208-217).

Ansoff, H. I. (1975). Managing strategic surprise by response to weak signals. *California management review*, *18*(2), 21-33.

Blei, D. M., Ng, A. Y., & Jordan, M. I. (2003). Latent dirichlet allocation. *Journal of machine Learning research*, *3*(Jan), 993-1022.

Blei, D. M., & Lafferty, J. D. (2006, June). Dynamic topic models. In *Proceedings of the 23rd international conference on Machine learning* (pp. 113-120).

Börner, K., Huang, W., Linnemeier, M., Duhon, R., Phillips, P., Ma, N., ... & Price, M. (2010). Rete-netzwerk-red: analyzing and visualizing scholarly networks using the Network Workbench Tool. *Scientometrics*, *83*(3), 863-876.

Boyack, K. W., & Klavans, R. (2014). Creation of a highly detailed, dynamic, global model and map of science. *Journal of the Association for Information Science and Technology*, *65*(4), 670-685.

Chen, C. (2006). CiteSpace II: Detecting and visualizing emerging trends and transient patterns in scientific literature. *Journal of the American Society for information Science and Technology*, *57*(3), 359-377.

Chen, C., Chen, Y., Horowitz, M., Hou, H., Liu, Z., & Pellegrino, D. (2009). Towards an explanatory and computational theory of scientific discovery. *Journal of Informetrics*, *3*(3), 191-209.

Chen, C., Hu, Z., Liu, S., & Tseng, H. (2012). Emerging trends in regenerative medicine: a scientometric analysis in CiteSpace. *Expert opinion on biological therapy*, *12*(5), 593-608.

Cozzens, S., Gatchair, S., Kang, J., Kim, K. S., Lee, H. J., Ordóñez, G., & Porter, A. (2010). Emerging technologies: quantitative identification and measurement. *Technology Analysis & Strategic Management*, *22*(3), 361-376.

Council of the European Union, 2024. Council recommendation on enhancing research security. Brussels.

De Moya-Anegon, F., Guerrero-Bote, V. P., López-Illescas, C., & Moed, H. F. (2018). Statistical relationships between corresponding authorship, international co-authorship and citation impact of national research systems. *Journal of informetrics*, *12*(4), 1251-1262.

Ebadi, A., Auger, A., & Gauthier, Y. (2026). WISDOM: An AI-powered framework for emerging research detection using weak signal analysis and advanced topic modelling. *Journal of Informetrics*, *20*(1), 101759.

European Commission (2023). Commission Recommendation of 3.10.2023 on critical technolgy areas for the EU's economic security for further risk assessment with Member States. Strasbourg, No. C(2023) 6689

European Commission and High Representative of the Union for Foreign Affairs and Security Policy (2023). Joint Communication to the European Parliament, the European Council and the Council on 'European Economic Security Strategy'. Brussels: European Commission, Communication No. JOIN(2023)20

Gazni, A., Sugimoto, C. R., & Didegah, F. (2012). Mapping world scientific collaboration: Authors, institutions, and countries. Journal of the American Society for Information Science and Technology, *63*(2), 323-335

Glanzel, W. (2001). National characteristics in international scientific co-authorship relations. *Scientometrics*, *51*(1), 69–115.

Glanzel, W., & Schubert, A. (2001). Double effort = double impact? A critical view at international co-authorship in chemistry. *Scientometrics*, *50*(2), 199–214.

Griffiths, T. L., & Steyvers, M. (2004). Finding scientific topics. *Proceedings of the National academy of Sciences*, *101*(suppl_1), 5228-5235.

Grootendorst, M. (2022). BERTopic: Neural topic modeling with a class-based TF-IDF procedure. arXiv, doi: 10.48550/arXiv.2203.05794

Hannas, W.C. and Tatlow, D.K., 2020. China's quest for foreign technology: beyond espionage. Oxon and New York: Routledge.

Holopainen, M., & Toivonen, M. (2012). Weak signals: Ansoff today. *Futures*, *44*(3), 198-205.

James, A., Flanagan, K., Naisbitt, A., & Rigby, J. (2025). EUROPEAN RESEARCH SECURITY: THREAT PERSPECTIVES AND THE RESPONSES OF POLICY MAKERS AND RESEARCH PERFORMING ORGANISATIONS.

Jiang, J., Ying, F., & Dhuny, R. (2025). Unveiling Technological Evolution with a Patent-Based Dynamic Topic Modeling Framework: A Case Study of Advanced 6G Technologies. *Applied Sciences*, *15*(7), 3783.

Kleinberg, J. (2002, July). Bursty and hierarchical structure in streams. In *Proceedings of the eighth ACM SIGKDD international conference on Knowledge discovery and data mining* (pp. 91-101).

Kölbl, L., & Grottke, M. (2020). Obtaining More Specific Topics and Detecting Weak Signals by Topic Word Selection. In *Reliability and Statistical Computing: Modeling, Methods and Applications* (pp. 193-206). Cham: Springer International Publishing.

Larivière, V., Gingras, Y., Sugimoto, C. R., & Tsou, A. (2015). Team size matters: Collaboration and scientific impact since 1900. *Journal of the Association for Information Science and Technology*, *66*(7), 1323-1332.

Lu, K., Yang, G., & Wang, X. (2022). Topics emerged in the biomedical field and their characteristics. *Technological Forecasting and Social Change*, *174*, 121218.

Lundin, H., & Shih, T. (2025). Governing the de-risking agenda: Policy instrument mixes in Nordic collaboration with China.

Ma, M., Mao, J., & Li, G. (2024). Discovering weak signals of emerging topics with a triple-dimensional framework. *Information Processing & Management*, *61*(5), 103793.

Matthews, D. (2025, November 11). *EU plans to ban Chinese universities from half of Horizon Europe*. *Science|Business*. Retrieved from https://sciencebusiness.net/news/r-d-funding/horizon-europe/eu-plans-ban-chinese-universities-half-horizon-europe

McInnes, L., Healy, J., & Melville, J. (2018). Umap: Uniform manifold approximation and projection for dimension reduction. *arXiv preprint arXiv:1802.03426*.

Mühlroth, C., & Grottke, M. (2018). A systematic literature review of mining weak signals and trends for corporate foresight. *Journal of Business Economics*, *88*(5), 643-687.

Ohniwa, R., Hibino, A., & Takeyasu, K. (2010). Trends in research foci in life science fields over the last 30 years monitored by emerging topics. *Scientometrics*, *85*(1), 111-127.

Park, C., & Kim, M. (2021). A study on the characteristics of academic topics related to renewable energy using the structural topic modeling and the weak signal concept. *Energies*, *14*(5), 1497.

Park, C., & Kim, M. (2024). Utilization and challenges of artificial intelligence in the energy sector. *Energy & Environment*, 0958305X241258795.

Pépin, L., Kuntz, P., Blanchard, J., Guillet, F., & Suignard, P. (2017). Visual analytics for exploring topic long-term evolution and detecting weak signals in company targeted tweets. *Computers & Industrial Engineering*, *112*, 450-458.

Persson, O., Glanzel, W., & Danell, R. (2004). Inflationary bibliometric values: The role of scientific collaboration and the need for relative indicators in evaluative studies. *Scientometrics*, *60*(3), 421–432.

Qiu, S., Steinwender, C., & Azoulay, P. (2025). Paper tiger? Chinese science and home bias in citations. *Journal of International Economics*, *157*, 104123.

Roche, I., Besagni, D., François, C., Hörlesberger, M., & Schiebel, E. (2010). Identification and characterisation of technological topics in the field of Molecular Biology. *Scientometrics*, *82*(3), 663-676.

Rotolo, D., Hicks, D., & Martin, B. R. (2015). What is an emerging technology?. *Research policy*, *44*(10), 1827-1843.

Rüland, A. L., Rüffin, N., Wang, R., & Mauduit, J. C. (2025). The implementation of research security policies in Germany: exploring policy narratives across governance levels. *European Security*, 1-23.

Saritas, O., & Smith, J. E. (2011). The big picture–trends, drivers, wild cards, discontinuities and weak signals. *Futures*, *43*(3), 292-312.

Shibata, N., Kajikawa, Y., Takeda, Y., & Matsushima, K. (2008). Detecting emerging research fronts based on topological measures in citation networks of scientific publications. *Technovation*, *28*(11), 758-775.

Schiebel, E., Hörlesberger, M., Roche, I., François, C., & Besagni, D. (2010). An advanced diffusion model to identify emergent research issues: the case of optoelectronic devices. *Scientometrics*, *83*(3), 765-781.

Shiu, S. H., & Lehti-Shiu, M. D. (2024). Assessing the evolution of research topics in a biological field using plant science as an example. *PLoS Biology*, *22*(5), e3002612.

Small, H., Boyack, K. W., & Klavans, R. (2014). Identifying emerging topics in science and technology. *Research policy*, *43*(8), 1450-1467.

The Economist, 2024. China has become a scientific superpower. London: The Economist.

Upham, S., & Small, H. (2010). Emerging research fronts in science and technology: patterns of new knowledge development. *Scientometrics*, *83*(1), 15-38.

Viet, N. T., & Gneushev, V. (2021, September). Analyzing and forecasting emerging technology trends by mining web news. In *Conference on Creativity in Intelligent Technologies and Data Science* (pp. 55-69). Cham: Springer International Publishing.

von der Leyen, U. 2023. Speech by President von der Leyen on EU-China relations to the Mercator Institute for China Studies and the European Policy Centre

Wang, J., Frietsch, R., Neuhäusler, P., & Hooi, R. (2024). International collaboration leading to high citations: Global impact or home country effect?. *Journal of Informetrics*, *18*(4), 101565.

Wieczorek, O., Unger, S., Riebling, J., Erhard, L., Koß, C., & Heiberger, R. (2021). Mapping the field of psychology: Trends in research topics 1995–2015. *Scientometrics*, *126*(12), 9699-9731.

Ye, G., Wang, C., Wu, C., Peng, Z., Wei, J., Song, X., Tan, Q., & Wu, L. (2023). Research frontier detection and analysis based on research grants information: A case study on health informatics in the US. Journal of Informetrics, 17(3), Article 101421.

Zhao, D., Tang, Z., & He, D. (2024). A systematic literature review of weak signal identification and evolution for corporate foresight. *Kybernetes*, *53*(10), 3160-3188.

## Appendix

**Table 1** Log-linear model coefficients for the effects of country and year. authorship type. and period on FNCI

| Coefficient | Estimate | Std Error | Statistic | p.value | Significance |
|---|---|---|---|---|---|
| (Intercept) | 0.618 | 0.009 | 69.007 | <0.001 | *** |
| EU-13 | -0.172 | 0.024 | -7.084 | <0.001 | *** |
| EU-14 | -0.001 | 0.015 | -0.094 | 0.925 | |
| Other | -0.02 | 0.012 | -1.716 | 0.086 | . |
| Russia | -0.335 | 0.031 | -10.839 | <0.001 | *** |
| Ukraine | -0.426 | 0.048 | -8.843 | <0.001 | *** |
| United States | -0.019 | 0.015 | -1.292 | 0.196 | |
| Unknown | -0.358 | 0.012 | -29.462 | <0.001 | *** |
| year | 0.015 | 0.003 | 4.333 | <0.001 | *** |
| EU-13:year | 0.007 | 0.01 | 0.728 | 0.466 | |
| EU-14:year | -0.014 | 0.006 | -2.426 | 0.015 | * |
| Other: year | -0.013 | 0.005 | -2.743 | 0.006 | ** |
| Russia: year | -0.022 | 0.013 | -1.621 | 0.105 | |
| Ukraine: year | 0.001 | 0.019 | 0.068 | 0.946 | |
| United States: year | -0.026 | 0.006 | -4.162 | <0.001 | *** |
| Unknown: year | -0.024 | 0.005 | -5.121 | <0.001 | *** |

**Table 2** Predicted FNCI from log-linear model by publication year and country

| Publication year | Country | Predicted FNCI | FNCI lower | FNCI upper |
|---|---|---|---|---|
| | China | 0.86 | 0.82 | 0.89 |
| | EU-13 | 0.56 | 0.50 | 0.63 |
| | EU-14 | 0.85 | 0.81 | 0.90 |
| | Other | 0.82 | 0.79 | 0.85 |
| | Russia | 0.33 | 0.25 | 0.41 |
| | Ukraine | 0.21 | 0.10 | 0.33 |
| | United States | 0.82 | 0.78 | 0.86 |
| 2020 | Unknown | 0.30 | 0.28 | 0.32 |
| | China | 0.97 | 0.94 | 1.00 |
| 2024 | EU-13 | 0.71 | 0.64 | 0.78 |
| | EU-14 | 0.86 | 0.82 | 0.90 |
| | Other | 0.84 | 0.81 | 0.86 |
| | Russia | 0.29 | 0.21 | 0.38 |
| | Ukraine | 0.29 | 0.19 | 0.40 |
| | United States | 0.75 | 0.70 | 0.79 |
| | Unknown | 0.25 | 0.23 | 0.27 |

**Table 3** Log-linear model coefficients for the effects of country. year. and signal type on FNCI

| Coefficient | **Estimate** | **Std Error** | **Statistic** | **p.value** | **Significance** |
|---|---|---|---|---|---|
| (Intercept) | 0.692 | 0.024 | 28.983 | <0.001 | *** |
| EU-13 | -0.131 | 0.066 | -1.985 | 0.047 | * |
| EU-14 | 0.064 | 0.04 | 1.606 | 0.108 | |
| Other | 0.039 | 0.032 | 1.219 | 0.223 | |
| Russia | -0.322 | 0.099 | -3.255 | 0.001 | ** |
| Ukraine | -0.634 | 0.158 | -4.02 | <0.001 | *** |
| United States | 0.054 | 0.042 | 1.293 | 0.196 | |
| Unknown | -0.382 | 0.032 | -12.089 | <0.001 | *** |
| year | 0.007 | 0.008 | 0.916 | 0.360 | |
| Weak-signal topics | 0.009 | 0.047 | 0.183 | 0.855 | |
| EU-13: year | -0.01 | 0.023 | -0.432 | 0.666 | |
| EU-14: year | -0.035 | 0.014 | -2.457 | 0.014 | * |
| Other: year | -0.038 | 0.011 | -3.385 | <0.001 | *** |
| Russia: year | -0.015 | 0.038 | -0.4 | 0.689 | |
| Ukraine: year | 0.091 | 0.053 | 1.724 | 0.085 | . |
| United States: year | -0.035 | 0.015 | -2.299 | 0.022 | * |
| Unknown: year | -0.02 | 0.01 | -1.97 | 0.049 | * |
| EU-13: weak-signal topics | -0.067 | 0.129 | -0.519 | 0.604 | |
| EU-14: weak-signal topics | -0.084 | 0.08 | -1.052 | 0.293 | |
| Other: weak-signal topics | -0.063 | 0.063 | -1.001 | 0.317 | |
| Russia: weak-signal topics | 0.102 | 0.173 | 0.59 | 0.555 | |
| Ukraine: weak-signal topics | 0.23 | 0.27 | 0.852 | 0.394 | |
| United States: weak-signal topics | -0.124 | 0.081 | -1.537 | 0.124 | |
| Unknown: weak-signal topics | -0.03 | 0.061 | -0.486 | 0.627 | |
| year: weak-signal topics | -0.005 | 0.015 | -0.309 | 0.757 | |
| EU-13: year: weak-signal topics | 0.023 | 0.045 | 0.513 | 0.608 | |
| EU-14: year: weak-signal topics | 0.042 | 0.028 | 1.49 | 0.136 | |
| Other: year: weak-signal topics | 0.027 | 0.021 | 1.288 | 0.198 | |
| Russia: year: weak-signal topics | -0.025 | 0.061 | -0.411 | 0.681 | |
| Ukraine: year: weak-signal topics | -0.106 | 0.093 | -1.142 | 0.254 | |
| United States: year: weak-signal topics | 0.029 | 0.029 | 0.979 | 0.328 | |
| Unknown: year: weak-signal topics | 0.001 | 0.019 | 0.049 | 0.961 | |

**Table 4** Predicted FNCI from log-linear model by publication year, topic type and country

| **Publication year** | **Topic type** | **Country** | **Predicted FNCI** | **FNCI lower** | **FNCI upper** |
|---|---|---|---|---|---|
| 2020 | Strong-signal topics | Unknown | 0.36 | 0.31 | 0.42 |
| | | Other | 1.08 | 0.99 | 1.16 |
| | | United States | 1.11 | 0.97 | 1.25 |
| | | EU-14 | 1.13 | 1.00 | 1.26 |
| | | China | 1.00 | 0.91 | 1.09 |
| | | Russia | 0.45 | 0.20 | 0.75 |
| | | EU-13 | 0.75 | 0.55 | 0.98 |
| | | Ukraine | 0.06 | -0.22 | 0.44 |
| | Weak-signal topics | Unknown | 0.33 | 0.25 | 0.42 |
| | | Other | 0.97 | 0.83 | 1.11 |
| | | United States | 0.88 | 0.68 | 1.10 |
| | | EU-14 | 0.97 | 0.77 | 1.20 |
| | | China | 1.01 | 0.86 | 1.18 |
| | | Russia | 0.62 | 0.24 | 1.11 |
| | | EU-13 | 0.65 | 0.35 | 1.02 |
| | | Ukraine | 0.34 | -0.12 | 1.05 |
| | | Unknown | 0.33 | 0.25 | 0.42 |
| 2024 | Strong-signal topics | Unknown | 0.29 | 0.26 | 0.33 |
| | | Other | 0.84 | 0.78 | 0.90 |
| | | United States | 0.89 | 0.78 | 1.00 |
| | | EU-14 | 0.91 | 0.81 | 1.01 |
| | | China | 1.06 | 0.99 | 1.12 |
| | | Russia | 0.40 | 0.17 | 0.67 |
| | | EU-13 | 0.74 | 0.59 | 0.90 |
| | | Ukraine | 0.57 | 0.26 | 0.96 |
| | Weak-signal topics | Unknown | 0.25 | 0.20 | 0.30 |
| | | Other | 0.91 | 0.81 | 1.01 |
| | | United States | 0.85 | 0.68 | 1.03 |
| | | EU-14 | 1.05 | 0.87 | 1.25 |
| | | China | 1.04 | 0.93 | 1.14 |
| | | Russia | 0.39 | 0.12 | 0.72 |
| | | EU-13 | 0.76 | 0.51 | 1.06 |